\documentclass[preprintnumbers,article,amsmath,amssymb,floatfix,10pt,prd,onecolumn,
superscriptaddress,nofootinbib]{revtex4}
\usepackage{bm}
\usepackage{amsfonts}
\usepackage{latexsym}
\usepackage[latin1]{inputenc}
\usepackage{graphicx}
\usepackage{amsmath}
\usepackage{palatino}
\usepackage{mathpazo}
\usepackage{textcomp}
\usepackage{float}
\usepackage{booktabs}
\usepackage{dcolumn}
\usepackage{ragged2e}
\usepackage{hyperref}
\hypersetup{colorlinks,citecolor=blue}
\hypersetup{colorlinks=true,linkcolor=red,filecolor=magenta,    urlcolor=blue}
\usepackage{amsmath}
\usepackage{xcolor}
\usepackage{orcidlink}
\usepackage{epsfig}
\usepackage[caption=false]{subfig}
\usepackage{commath}
\def\jnl@style{\it}
\def\aaref@jnl#1{{\jnl@style#1}}

\def\aaref@jnl#1{{\jnl@style#1}}

\def\aj{\aaref@jnl{AJ}}                   
\def\apj{\aaref@jnl{ApJ}}                 
\def\apjl{\aaref@jnl{ApJ}}                
\def\apjs{\aaref@jnl{ApJS}}               
\def\apss{\aaref@jnl{Ap\&SS}}             
\def\aap{\aaref@jnl{A\&A}}                
\def\aapr{\aaref@jnl{A\&A~Rev.}}          
\def\aaps{\aaref@jnl{A\&AS}}              
\def\mnras{\aaref@jnl{Mon.~Not.~Roy.~Astron.~Soc.}}             
\def\prd{\aaref@jnl{Phys.~Rev.~D}}        
\def\prc{\aaref@jnl{Phys.~Rev.~C}}  
\def\prl{\aaref@jnl{Phys.~Rev.~Lett.}}    
\def\qjras{\aaref@jnl{QJRAS}}             
\def\skytel{\aaref@jnl{S\&T}}             
\def\ssr{\aaref@jnl{Space~Sci.~Rev.}}     
\def\zap{\aaref@jnl{ZAp}}                 
\def\nat{\aaref@jnl{Nature}}              
\def\aplett{\aaref@jnl{Astrophys.~Lett.}} 
\def\apspr{\aaref@jnl{Astrophys.~Space~Phys.~Res.}} 
\def\physrep{\aaref@jnl{Phys.~Rep.}}      
\def\physscr{\aaref@jnl{Phys.~Scr}}       
\def\commat{\aaref@jnl{Comm.~Math.~Phys.}}              
\def\science{\aaref@jnl{Science}}               
\def\cqg{\aaref@jnl{Classical Quant.~Grav.}}            
\def\jpcs{\aaref@jnl{JPCS}}                                     
\def\ijmpd{\aaref@jnl{Int.~J.~Mod.~Phys.~D}}                    
\def\grg{\aaref@jnl{Gen.~Relat.~Gravit.}}               
\def\rpp{\aaref@jnl{Rep.~Prog.~Phys.}}          
\def\npa{\aaref@jnl{Nucl.~Phys.~A}}        
\def\lrr{\aaref@jnl{Living Rev.~Rel.}}                   
\def\jcap{\aaref@jnl{J.~Cosmology Astropart.~Phys.}}    
\def\rmp{\aaref@jnl{Rev.~Mod.~Phys.}}   
\def\epjc{\aaref@jnl{Eur.~Phys.~J.~C}}
\def\plb{\aaref@jnl{~Phy.~Lett.~B}}
\def\mpla{\aaref@jnl{Mod.~Phy.~Lett.~A}}
\def\arxiv{\aaref@jnl{arxiv.org}}

\allowdisplaybreaks[1]

\begin{document}
\color{black}       
\title{Traversable wormhole inspired by non-commutative geometries in $f(Q)$ gravity with conformal symmetry}
\author{G. Mustafa\orcidlink{0000-0003-1409-2009}}
\email{gmustafa3828@gmail.com}
\affiliation{Department of Physics, Zhejiang Normal University,
Jinhua, 321004, People's Republic of China.}

\author{Zinnat Hassan\orcidlink{0000-0002-6608-2075}}
\email{zinnathassan980@gmail.com}
\affiliation{Department of Mathematics, Birla Institute of Technology and
Science-Pilani,\\ Hyderabad Campus, Hyderabad-500078, India.}

\author{P.K. Sahoo\orcidlink{0000-0003-2130-8832}}
\email{pksahoo@hyderabad.bits-pilani.ac.in}
\affiliation{Department of Mathematics, Birla Institute of Technology and
Science-Pilani,\\ Hyderabad Campus, Hyderabad-500078, India.}

\date{\today}
\begin{abstract}

This article is based on the study of wormhole geometries in the context of symmetric teleparallel gravity or $f(Q)$ gravity, where $Q$ is the non-metricity scalar, and it is responsible for the gravitational interaction. To discuss the wormhole solutions, we consider spherically symmetric static spacetime metric with anisotropic matter contents under well-known non-commutative distributions known as Gaussian and Lorentzian distributions with an extra condition of permitting conformal killing vectors (CKV). This work aims to obtain wormhole solutions under these distributions, and through we found that wormhole solutions exist under these Gaussian and Lorentzian sources with viable physical properties. Further, we examine the stability of our obtained solutions through Tolman-Oppenheimer-Volkoff (TOV) equation and found that our calculated results are stable.
\end{abstract}
\maketitle

\date{\today}

\section{Introduction}

In recent decades, wormhole physics has been of considerable interest among researchers after the seminal work done by Morris and Throne \cite{Morris/1988}. The possibility of the existence of wormholes is still an open question. Khatsymovsky discussed deeply on this topic in \cite{Khatsymovsky/1994}. Gravitational lensing by wormholes was studied in \cite{Jusufi/2018,Javed/2019,Ali/2019} and light deflection in \cite{Ali/2018,Ali/2020}. Recently, some interesting investigations on wormhole geometries have been discussed in \cite{Halilsoy/2014,Richarte/2017,Maeda,Samart}.\\
Basically, a wormhole is a hypothetical tunnel in space-time with two distinct endpoints or two connecting black holes, which was first introduced by Weyl \cite{Coleman/1985} and, after that, by Wheeler \cite{Wheeler/1957}. Morris and Throne\cite{Thorne/1988} and other authors in \cite{Ellis/1973,Bronnikov/1973,Clement/1984} argued that a wormhole would permit travel in space and time and explicitly discussed the best way to convert a wormhole traversing space into traversing time. Moreover, some literature is available on wormholes where the traversing path does not go through the exotic matter region \cite{Visser/1996,Visser/1989}.\\
Topologically, in the space-time geometry, wormholes act like tunnels that connect two distinct space-times of the universe by a minimal surface called the wormhole throat and satisfy the flaring out conditions \cite{Hochberg/1997}, through which one can easily traverse in both dimensions. Recent research is based on the study of the essential conditions to guarantee their traversability. The prominent of these properties is a unique type of matter called exotic matter, and it is responsible for the violation of null energy conditions (NEC) and is necessary for forming a traversable wormhole. Such strange object supported by a single fluid component exist on both dynamic \cite{Dehghani/2009,Bochicchio/2010,DeBenedictis/2008,Gonzalez/2003,Cataldo/2011,Hansen/2009} and static \cite{Anabalon/2012,Balakin/2010,Jamil/2010,Cataldo/2002} cases. In general, classical matters obey the energy conditions, but some quantum fields like the Casimir effect, scalar-tensor theories, and Hawking evaporation supported the violations of energy conditions. Visser et al. \cite{Dadhich/2003} developed a suitable measure for wormhole maintenance called volume integral quantifier for quantifying the total averaged null energy condition.\\
Nevertheless, in this context of Einstein General relativity (GR), a static wormhole without violating the energy conditions is still an open question. Hence, Visser \cite{M. Visser/1989,Visser/1989a} introduced a technique called the cut and paste technique to minimize the usage of exotic matter, but it restricted the exotic fluid at the throat of the wormhole. Also, Kuhfittig \cite{Kuhfittig/2002,Kuhfittig/1999} developed a solution to hamper the exotic fluid of an arbitrary thin region by forcing a condition $b^{'}(r)<1$ at wormhole's throat.\\
As it is known that modified gravity theories might significantly minimize or even invalidate the requirement for exotic matter. In fact, the investigation of wormhole solutions in different modified theories of gravity is significant in Theoretical Physics. S. N. Lobo and Oliveira \cite{Oliveira/2009} have constructed traversable wormhole geometries in $f(R)$ gravity where $R$ represents the Ricci scalar, and by using specific shape function and several equations of state, they investigated the validity of energy conditions. DeBenedictis and Horvat have studied the existence of wormhole throats for $R^n$ model under anisotropic fluid in $f(R)$ gravity \cite{DeBenedictis/2012}. Harko $et\, al.$ \cite{Harko/2013} and Pavlovic and Sossich \cite{Pavlovic/2015} have investigated the wormhole geometries without exotic matter in modified $f(R)$ gravity. Furthermore, one can check wormhole solutions in  various modified theories of gravity such as in $f(R,T)$ gravity \cite{Moraes/2019,Sahoo/2018}, $f(T)$ gravity \cite{Sharif/2013,Sharif/2014,Sharif/2013a}, noncommutative geometry \cite{Rahaman/2012,Bhar/2014,Hussain/2021} and so on.\\
Basically, non-commutative geometry is an intrinsic property of the manifold itself as explicit in \cite{Smailagic/2006}, and it can be introduced in GR by modifying the matter source. Schneider and DeBenedictis discussed in detail the background of both non-commutative distributions in \cite{Mathew/2020}. It is accepted that with the assistance of non-commutative geometries, some viewpoints of quantum gravity can be investigated mathematically in a preferable way. Spacetime quantization is the essential development of string theory, and the spacetime coordinates may be operated as non-commutative operators in the D-brane \cite{Seiberg}. Such non-commutative operator are used to encrypted in the commutator $[x^{\mu},x^{\nu}]=i\theta^{\mu\nu}$, where $\theta^{\mu\nu}$ is the anti-symmetric matrix which indicates the discretization of spacetime \cite{Doplicher/1994,Kase/2003,Smailagic/2004,Nicolini/2009}. Non-commutativity replaces the point-like structures with smeared objects to wipe out the divergences that show up in GR. This smearing can be shown by utilizing Gaussian and Lorentzian distributions of minimal length $\sqrt{\theta}$ rather than the Dirac delta function. Sushkov \cite{Sushkov/2005} employed Gaussian distribution to study wormholes supported by phantom energy. Rahaman et al. \cite{Islam/2012} studied wormhole solutions under non-commutative background and found that wormhole solutions exist in the usual four, as well as in five dimensions, but they do not exist in higher-dimensional spacetimes. Also, the stability of a special class of thin-shell wormholes in GR under non-commutative geometry has been studied in \cite{Kuhfittig/2012a}.\\
The study of conformal symmetry defines a link between matter and geometry through the Einstein field equations.  Due to this, the vector $\varrho$ defined as the generator and the metric $g$ is mapped onto itself conformally along $\xi$ of this conformal symmetry, and it can be written as
\begin{equation}\label{a}
\mathfrak{L}_{\varrho}g_{\zeta\eta}=F(r)g_{\zeta\eta},
\end{equation}
where $F(r)$ represents the conformal killing vector (CKV) and $\mathfrak{L}$ represents the Lie derivative operator. An interesting observation found that for a static metric, neither $\varrho$ nor $F(r)$ need to be static. The authors in \cite{Herrera/1984,Herrera/1985} used this approach to show that for a class of conformal motions, the equation of state (EoS) is uniquely determined by the Einstein equations. Later on, Maartens and Maharaj \cite{Maartens/1990} has extended this particular exact solution for the static sphere of charge imperfect fluid. Recently, Kuhfitting \cite{Kuhfittig/2016a,Kuhfittig/2015a} discussed a barotropic equation of state admitting a one-parameter group of conformal motion for the wormhole.\\
Recently, an interesting alternative theory of gravity has been proposed by Jimenez et al. \cite{Jimenez/2018}, so-called $f(Q)$ gravity or symmetric teleparallel gravity where $Q$ represents the non-metricity scalar. The critical difference between GR classical gravity and symmetric teleparallel gravity is the affine connection rather than the physical manifold. In \cite{Jimenez/2018}, the authors have shown that symmetric teleparallel gravity is equivalent to GR in flat space. Noted that $f(Q)$ gravity features in second-order field equations, which is similar to $f(T)$ gravity. While $f(R)$ gravity field equations are up to fourth-order \cite{T.P.}. Thus, $f(Q)$ gravity presents a distinct geometric description of gravity. Since it is newly proposed, many authors have already made some interesting applications of this gravity. We quote, for instance, in \cite{Frusciante/2021,Heisenberg/2020,Bajardi/2020} some cosmological features of $f(Q)$ gravity were investigated, Energy conditions in $f(Q)$ gravity have been studied by Mandal et al. in \cite{Mandal/2020} and wormhole solutions have been investigated in $f(Q)$ gravity in Refs. \cite{Zinnat/2021,Hassan/2021}. For more application of $f(Q)$ gravity one can check \cite{Lazkoz/2019,Barrosa/2020}.\\
In this present study, we are inspired to discuss the existence of wormhole solutions under Gaussian and Lorentzian distribution with conformal symmetry in $f(Q)$ gravity. The outline of the manuscript is organized as follows: In sec. \ref{sec2}, we present the necessary formulation for $f(Q)$ gravity. The basic condition for traversable wormhole and the outline of conformal killing vectors have been presented in sec. \ref{sec3}. In sec. \ref{sec4} we investigate the wormhole solutions inspired by Gaussian and Lorentzian sources. Moreover, the stability of the obtain wormhole solutions with the help of Tolman-Oppenheimer-Volkoff has been discussed in sec. \ref{sec5} while in sec. \ref{sec6} we use the exoticity parameter to investigate the exotic matter at the throat of the wormhole. At last, the concluding remarks of our current study are summarized in sec. \ref{sec7}.

\section{Basic Field Equations in $f(Q)$ gravity}\label{sec2}

The dynamics of the universe in symmetric teleparallel gravity is described by following action \cite{Jimenez/2018}
\begin{equation}\label{1}
\mathcal{S}=\int\frac{1}{2}\,f(Q)\sqrt{-g}\,d^4x+\int \mathcal{L}_m\,\sqrt{-g}\,d^4x\, ,
\end{equation}
where $f(Q)$ represents the function of the non-metricity scalar $Q$, $g$ denote the determinant of the metric $g_{\mu\nu}$, and $\mathcal{L}_m$ is the density of matter Lagrangian.\\
Another essential component to describe the $f(Q)$ gravity is the non-metricity tensor which is defined as\\
\begin{equation}\label{2}
Q_{\lambda\mu\nu}=\bigtriangledown_{\lambda} g_{\mu\nu},
\end{equation}
and its two traces are given below
\begin{equation}
\label{3}
Q_{\alpha}=Q_{\alpha}\;^{\mu}\;_{\mu},\; \tilde{Q}_\alpha=Q^\mu\;_{\alpha\mu}.
\end{equation}
Further, the conjugate of non-metricity tensor (superpotential tensor) is of the form:
\begin{equation}\label{4}
P^\alpha\;_{\mu\nu}=\frac{1}{4}\left[-Q^\alpha\;_{\mu\nu}+2Q_{(\mu}\;^\alpha\;_{\nu)}+Q^\alpha g_{\mu\nu}-\tilde{Q}^\alpha g_{\mu\nu}-\delta^\alpha_{(\mu}Q_{\nu)}\right],
\end{equation}
Now one can find the non-metricity scalar by taking the trace of non-metricity tensor given by Eq. \eqref{2}
\begin{equation}
\label{5}
Q=-Q_{\alpha\mu\nu}\,P^{\alpha\mu\nu}.
\end{equation}
The definition of the stress-energy tensor for the fluid description of the spacetime is of the form
\begin{equation}\label{6}
T_{\mu\nu}=-\frac{2}{\sqrt{-g}}\frac{\delta\left(\sqrt{-g}\,\mathcal{L}_m\right)}{\delta g^{\mu\nu}}.
\end{equation}
The field equation describing the gravitational interactions in the symmetric teleparallel gravity acquired by varying the action \eqref{1} with respect to the fundamental metric tensor $g_{\mu\nu}$ is shown below:
\begin{equation}\label{7}
\frac{2}{\sqrt{-g}}\bigtriangledown_\gamma\left(\sqrt{-g}\,f_Q\,P^\gamma\;_{\mu\nu}\right)+\frac{1}{2}g_{\mu\nu}f \\
+f_Q\left(P_{\mu\gamma i}\,Q_\nu\;^{\gamma i}-2\,Q_{\gamma i \mu}\,P^{\gamma i}\;_\nu\right)=-T_{\mu\nu},
\end{equation}
where $f_Q=\frac{df}{dQ}$.\\
Also, by varying the action \eqref{1} with respect to the connection, one can find the following relation
\begin{equation}\label{8}
\bigtriangledown_\mu \bigtriangledown_\nu \left(\sqrt{-g}\,f_Q\,P^\gamma\;_{\mu\nu}\right)=0.
\end{equation}
\section{Traversability conditions for wormholes and Conformal killing vector}
\label{sec3}
Here, we consider spherically symmetric static spacetime metric and this spacetime is conventionally composed as
\begin{equation}\label{9}
 d{s}^2=-e^{\epsilon(r)}dt^2+e^{\sigma(r)}d{r}^2+r^2(d\theta^2+\sin^2 \theta d\Phi^2),
\end{equation}
where
\begin{itemize}
  \item $\epsilon(r)=2\,\Psi(r)$ with $\Psi(r)$ is the redshift function.
  \item $e^{\sigma(r)}=\left(\frac{r-b(r)}{r}\right)^{-1}$ with $b(r)$ is the shape function.
  \item The wormhole throat joins two asymptotic regions and is placed at the radial coordinate $r_0$, where $b(r_0)= r_0$.
  \item The flaring-out requirement,$\frac{b(r) - rb'(r)}{
2b^2(r)} > 0$, which should valid at or near the throat, must be satisfied by the shape function $b(r)$. This reduces to $b'(r_0)<1$ near the wormhole's throat.
\item The shape function should meet the condition $1-\frac{b(r)}{r}>0$ for the radial coordinates $r>r_0$ in order to maintain the proper signature of the metric.
\item The metric functions must obey the requirements $\Psi(r)$ and $\frac{b(r)}{r}$ tend to zero as r approaches to $\infty$ in order to have asymptotically flat geometries. For no-asymptotically flat wormholes, these criteria can obviously be relaxed.
\end{itemize}
In the present study, as we are going to analyze the wormhole solutions, we assume the matter content described by an anisotropic energy-momentum tensor which is given by
\begin{equation}\label{10}
T_{\mu}^{\nu}=\left(\rho+p_t\right)u_{\mu}\,u^{\nu}+p_t\,\delta_{\mu}^{\nu}+\left(p_r-p_t\right)v_{\mu}\,v^{\nu},
\end{equation}
where $\rho$ is the energy density and  $v_{\mu}$ is the unitary space-like vector in the radial direction. $u_{\mu}$ is the four-velocity vector such that $-u^{\mu}u_{\mu}=v_{\mu}v^{\mu}=1$. The expressions $p_r$ and $p_t$ denotes the radial and tangential pressures, respectively and both are functions of radial coordinate $r$. \\
For the metric \eqref{9}, the trace of the non-metricity  $Q$ is given by,
\begin{equation}\label{11}
Q=\frac{2}{r}e^{-\sigma(r)}\left(\epsilon^{'}(r)+\frac{1}{r}\right).
\end{equation}
Now, in our current work, we assume a linear functional form of $f(Q)$ gravity, which is expressed as:
\begin{equation}\label{12}
f(Q)=\alpha Q+\beta
\end{equation}
where, $\alpha$ and $\beta$ are free parameters.\\

Further, we use the concept of conformal symmetry by using the vector field $F$, the Eq. \eqref{a} provides the following expression:
\begin{equation}\label{13}
\mathfrak{L}_{\varrho}g_{\zeta\eta}=g_{\eta\eta}\varrho^{\eta}_{;\lambda}+g_{\zeta\eta}\varrho^{\eta}_{;\eta}=F(r)g_{\zeta\eta}.
\end{equation}
where $\mathfrak{L}$ represents the Lie derivative, with the CKVs $\varrho^{\eta}$, and vector field $F(r)$. With the help of killing vector, constants of the motion can be determined i.e. along any given geodesic, quantities will be constant. Moreover, for (a) $F(r)=constant$, Eq. \eqref{13} gives Homothetic vector, for (b) $F(r)=0$, Eq. \eqref{13} gives the killing vector, and when (c) $F(r)=F(x,t)$ then it gives conformal vector.

Using Eq.(\ref{9}), in Eq.(\ref{13}), we get the following three different expressions:
\begin{eqnarray*}
\varrho^{1}\zeta^{'}(r)=F(r),\;\;\;\;\;\;
\varrho^{1}=\frac{r F(r)}{2},\;\;\;\;\;\;
\varrho^{1}\eta^{'}(r)+2\varrho^{1}_{,1}=F(r).
\end{eqnarray*}
By solving the above system by using the spacetime Eq. (\ref{9}), we get the following relations:
\begin{equation}
e^{\epsilon (r)}=e^{2\Psi(r)}=K_{1}^{2}r^2, \;\;\;\;\;\;\;\;e^{\sigma (r)}=\left(\frac{r-b(r)}{r}\right)^{-1}=\frac{K^{2} _{2}}{(F(r))^{2}},\label{14}
\end{equation}
where, $K_1$ and $K_2$ are considered integration constants. By using Eqs. (\ref{9}-\ref{12}) with Eq. (\ref{14}) in Eq. (\ref{7}) we get a following final version of field equations as:
\begin{equation}
\label{15}
\rho=-\frac{\beta}{2}+\frac{\alpha}{r^2}-\frac{\alpha\,F(r)}{K_2^2\,r^2}\left(2\,r\,F^{'}(r)+F(r)\right),
\end{equation}
\begin{equation}
\label{16}
p_r=\frac{\beta }{2}+\frac{3 \alpha  (F(r))^2}{K^2_2 r^2}-\frac{\alpha }{r^2},
\end{equation}
\begin{equation}
\label{17}
p_t=\frac{\beta}{2}+\frac{\alpha\,F(r)}{K_2^2\,r^2}\left(2\,r\,F^{'}(r)+F(r)\right),
\end{equation}
where $'$ represents $\frac{d}{dr}$.

\section{Physical Analysis of wormhole geometry}
\label{sec4}

In this section, we shall discuss the physical analysis of wormhole solutions with the help of Eqs. \eqref{15}-\eqref{17} under noncommutative distributions. For this purpose, we consider two noncommutative sources, namely Gaussian and Lorentzian distribution of particle-like gravitational sources, and hence the energy densities  are expressed as \cite{Smailagic,Spalluci}
\begin{equation}
\label{4a}
\rho=\frac{M e^{-\frac{r^2}{4 \theta }}}{8 \pi ^{3/2} \theta ^{3/2}},
\end{equation}
\begin{equation}
\label{4b}
\rho=\frac{\sqrt{\theta } M}{\pi ^2 \left(\theta +r^2\right)^2},
\end{equation}
respectively. Here, $M$ and $\theta$ are the smearing mass distribution and the noncommutativity parameter, respectively.

\subsection{Wormhole solutions under Gaussian Distribution for $f(Q)$ gravity}
\label{sec4I}

In this subsection, we match the Gaussian source of energy density \eqref{4a} with Eq. (\ref{15}), we get the following differential equation
\begin{equation}
\frac{M e^{-\frac{r^2}{4 \theta }}}{8 \pi ^{3/2} \theta ^{3/2}}=-\frac{\beta}{2}+\frac{\alpha}{r^2}-\frac{\alpha\,F(r)}{K_2^2\,r^2}\left(2\,r\,F^{'}(r)+F(r)\right),\label{19}
\end{equation} 
After solving the above differential equation with the scope of Eq. \eqref{13}, we get the following shape function
\begin{equation}
\frac{b}{\sqrt{\theta }}\left(\frac{r}{\sqrt{\theta }}\right)=-\frac{C_1}{K_2^2}-\frac{K^{2} _{2}}{8\,\alpha }\left(-\frac{2 M_1 erf\left(\frac{r}{2 \sqrt{\theta }}\right)}{\pi }+\frac{2 M_1 r e^{-\frac{1}{4} \left(\frac{r}{\sqrt{\theta }}\right)^2}}{\pi ^{3/2} \sqrt{\theta }}+\frac{8 \alpha  r}{\sqrt{\theta }}-\frac{1}{3} 4 \beta_{1} \left(\frac{r}{\sqrt{\theta }}\right)^3\right)+\frac{r}{\sqrt{\theta }},\label{20}
\end{equation} 
where $M_1=\frac{M}{\sqrt{\theta}}$, $\beta_{1}=\sqrt{\theta }\beta$, $C_{1}=\sqrt{\theta }C_{0}$ with $C_{0}$ is a constant of integration for Eq. (\ref{19}), and $erf$ is an error function.  By using the Eq. (\ref{20}) in Eq. (\ref{16}) and Eq. (\ref{17}), we get a following expressions for pressure components
\begin{eqnarray}
p_r&=&\frac{1}{\theta}\bigg(\frac{\beta_{1}}{3}-\frac{5 \alpha  C_{1}}{K^{4} _{2} \left(\frac{r}{\sqrt{\theta }}\right)^3}-\frac{5 M_1 erf\left(\frac{r}{2 \sqrt{\theta }}\right)}{4 \pi  \left(\frac{r}{\sqrt{\theta }}\right)^3}+\frac{\frac{5 M_1 e^{-\frac{1}{4} \left(\frac{r}{\sqrt{\theta }}\right)^2}}{4 \pi ^{3/2}}-4 \alpha }{\left(\frac{r}{\sqrt{\theta }}\right)^2}\bigg),\label{21}\\
p_t&=&\frac{1}{\theta}\bigg(\frac{2 \beta_{1}}{3}-\frac{5 \alpha  C_{1}}{2 K^{4} _{2} \left(\frac{r}{\sqrt{\theta }}\right)^3}-\frac{5 M_1 erf\left(\frac{r}{2 \sqrt{\theta }}\right)}{8 \pi  \left(\frac{r}{\sqrt{\theta }}\right)^3}-\frac{3 M_1 e^{-\frac{1}{4} \left(\frac{r}{\sqrt{\theta }}\right)^2}}{16 \pi ^{3/2}}+\frac{5 M_1 e^{-\frac{1}{4} \left(\frac{r}{\sqrt{\theta }}\right)^2}}{8 \pi ^{3/2} \left(\frac{r}{\sqrt{\theta }}\right)^2}-\frac{4 \alpha }{\left(\frac{r}{\sqrt{\theta }}\right)^2}\bigg)\label{22}.
\end{eqnarray}
\begin{figure}
\centering \epsfig{file=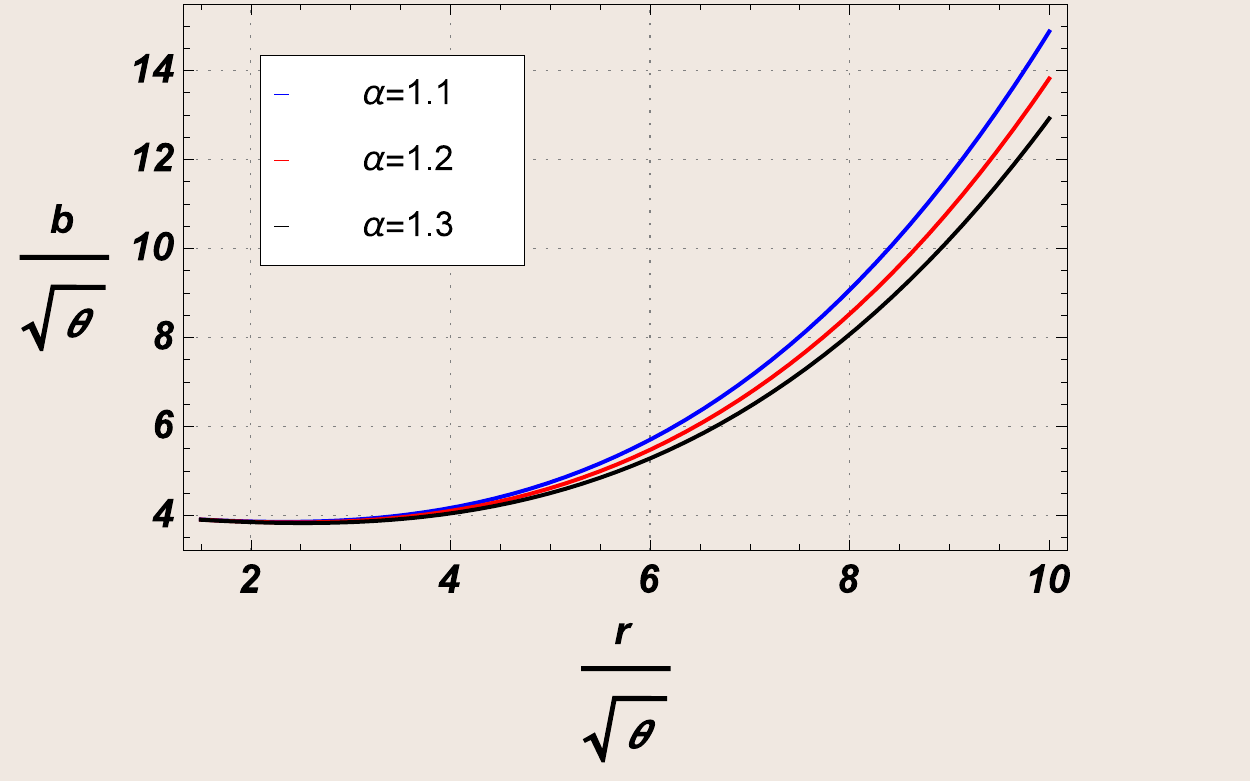, width=.48\linewidth,
height=2.5in}\epsfig{file=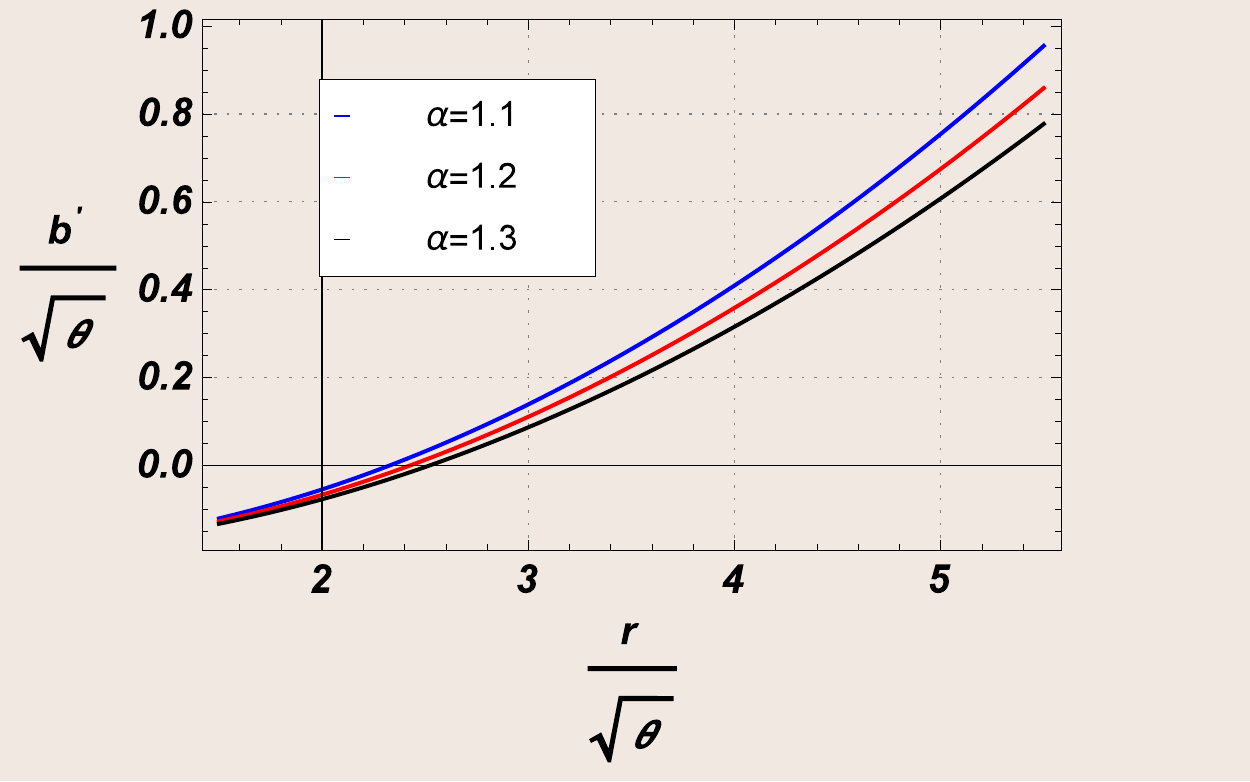, width=.48\linewidth,
height=2.5in}\caption{\label{Fig.1} Shows the behavior of $\frac{b}{\sqrt{\theta}}\left(\frac{r}{\sqrt{\theta}}\right)$ and $\frac{b^{'}}{\sqrt{\theta}}\left(\frac{r}{\sqrt{\theta}}\right)$ with respect to $\frac{r}{\sqrt{\theta}}$ for different values of $\alpha$ with fixed parameters $M_1=0.2$, $\beta_1=0.07$, $K_2=1.2$ and $C_1=-5$ under Gaussian distribution.}
\end{figure}
\begin{figure}
\centering \epsfig{file=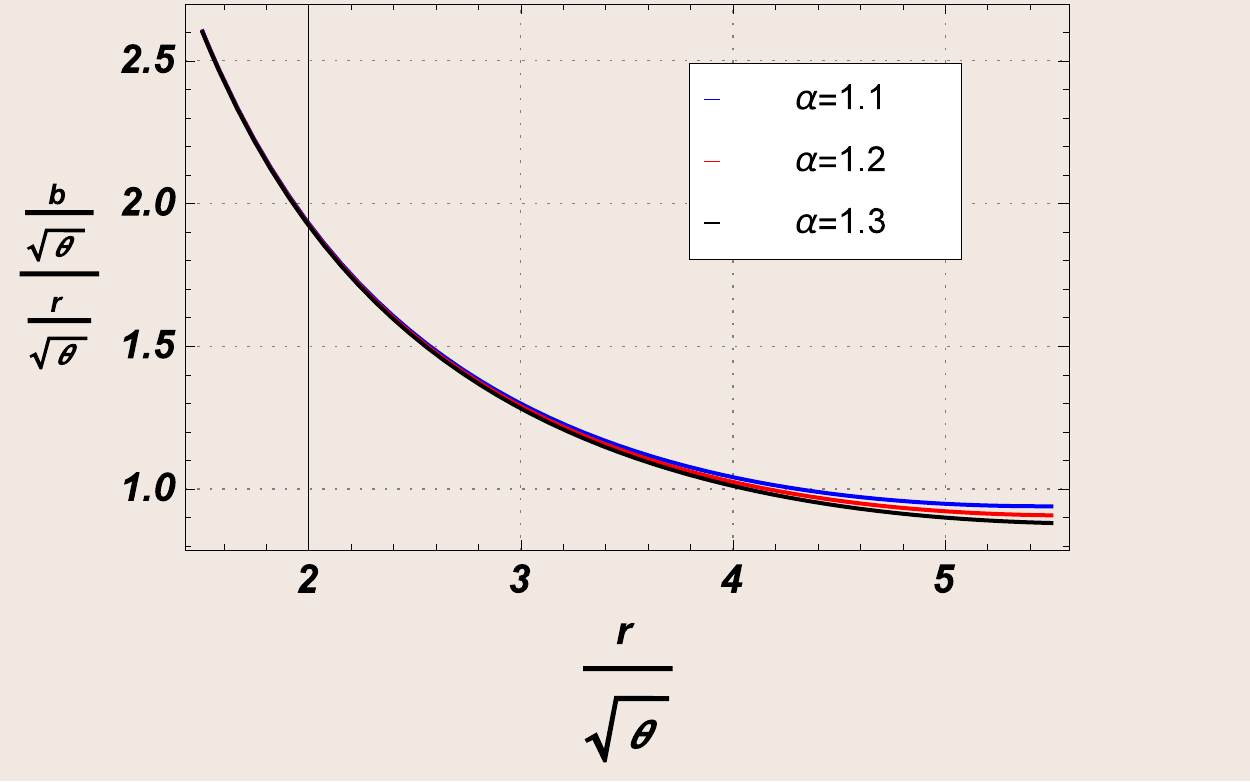, width=.48\linewidth,
height=2.5in}\epsfig{file=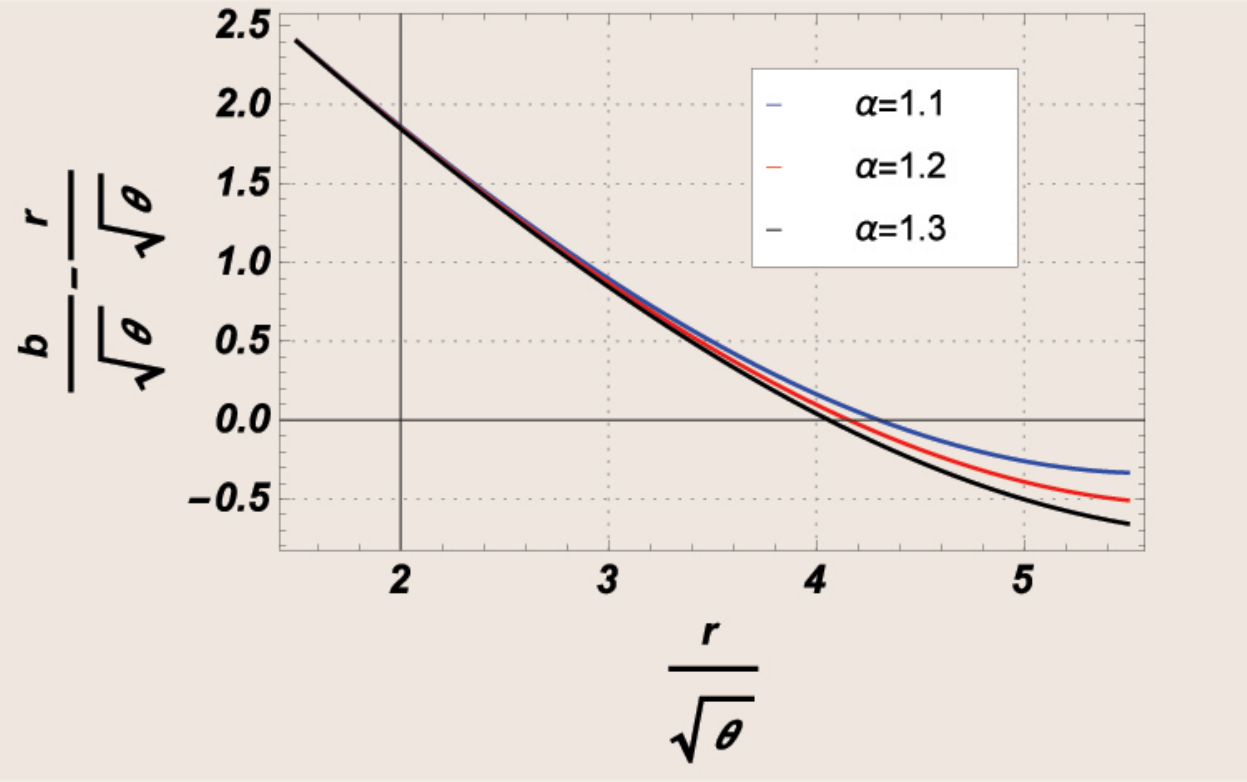, width=.48\linewidth,
height=2.5in}\caption{\label{Fig.2} Shows the behavior of $\frac{\frac{b}{\sqrt{\theta}}\left(\frac{r}{\sqrt{\theta}}\right)}{\frac{r}{\sqrt{\theta}}}$ and $\frac{b}{\sqrt{\theta}}\left(\frac{r}{\sqrt{\theta}}\right)-\left(\frac{r_0}{\sqrt{\theta}}\right)$ with respect to $\frac{r}{\sqrt{\theta}}$ for different values of $\alpha$ with fixed parameters $M_1=0.2$, $\beta_1=0.07$, $K_2=1.2$ and $C_1=-5$ under Gaussian distribution.}
\end{figure}
\begin{figure}
\centering \epsfig{file=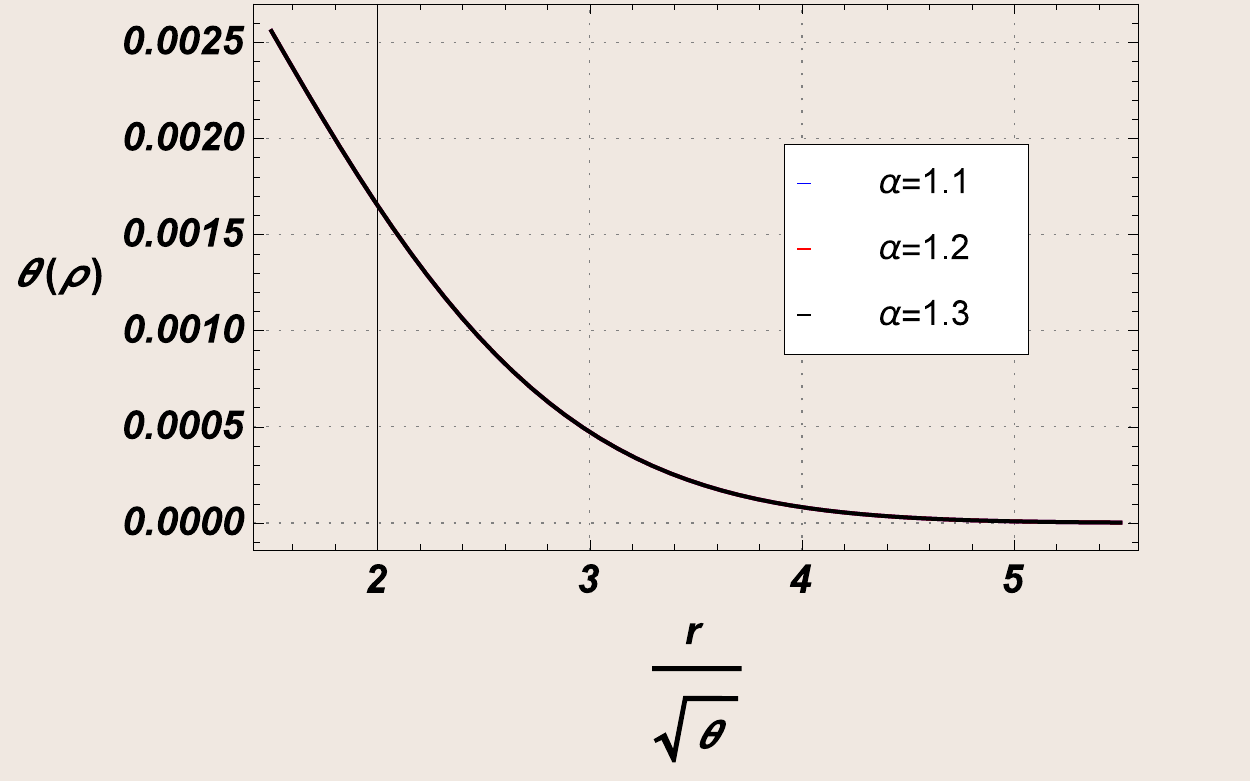, width=.48\linewidth,
height=2.5in}\caption{\label{Fig.3} Evolution of $\theta(\rho)$ with respect to $\frac{r}{\sqrt{\theta}}$ for different values of $\alpha$ with fixed parameters $M_1=0.2$, $\beta_1=0.07$, $K_2=1.2$ and $C_1=-5$ under Gaussian distribution.}
\end{figure}
\begin{figure}
\centering \epsfig{file=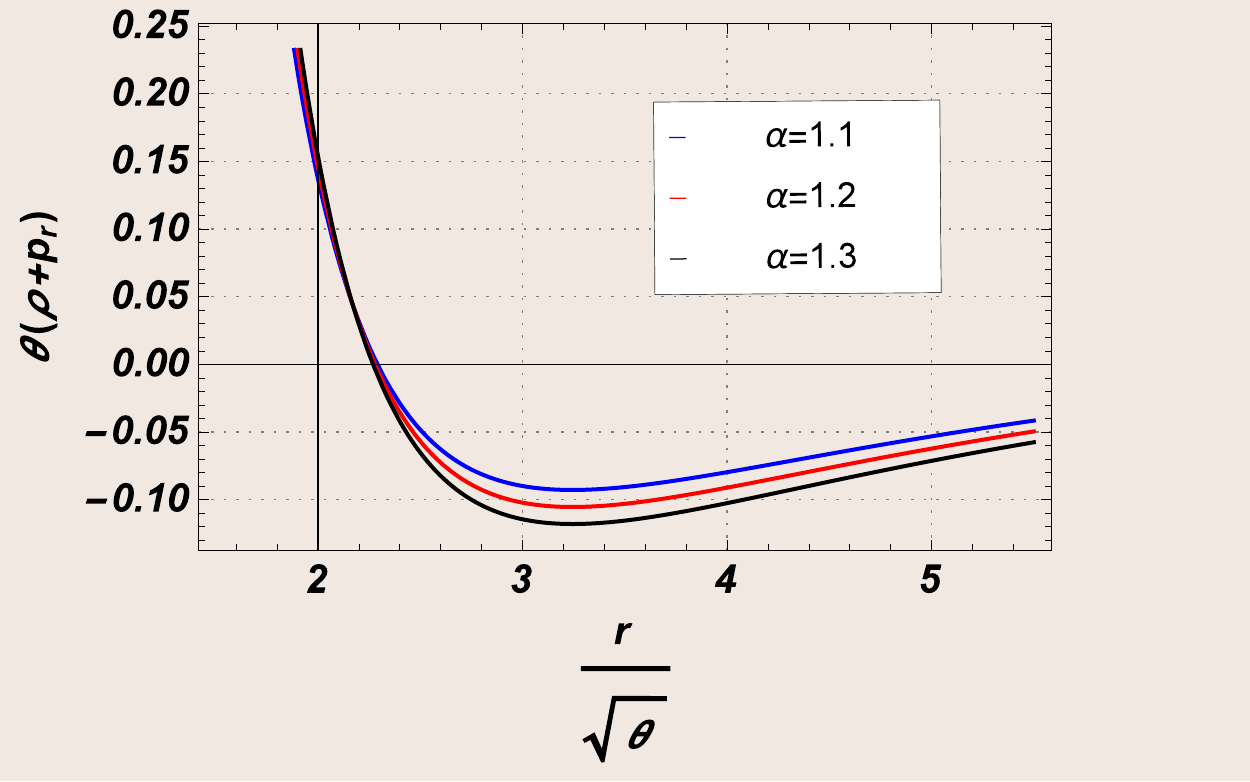, width=.48\linewidth,
height=2.5in}\epsfig{file=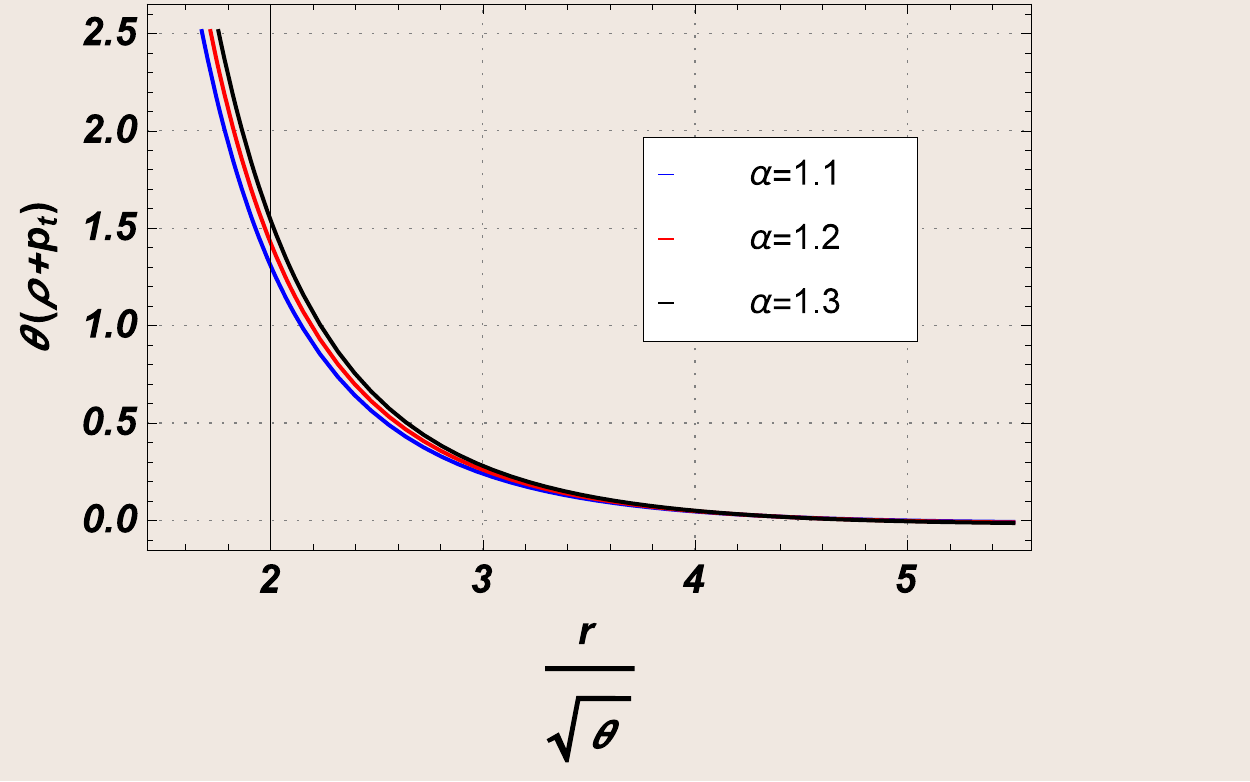, width=.48\linewidth,
height=2.5in}\caption{\label{Fig.4} Evolution of $\theta\left(\rho+p_r\right)$ and $\theta\left(\rho+p_t\right)$ with respect to $\frac{r}{\sqrt{\theta}}$ for different values of $\alpha$ with fixed parameters $M_1=0.2$, $\beta_1=0.07$, $K_2=1.2$ and $C_1=-5$ under Gaussian distribution.}
\end{figure}
\begin{figure}
\centering \epsfig{file=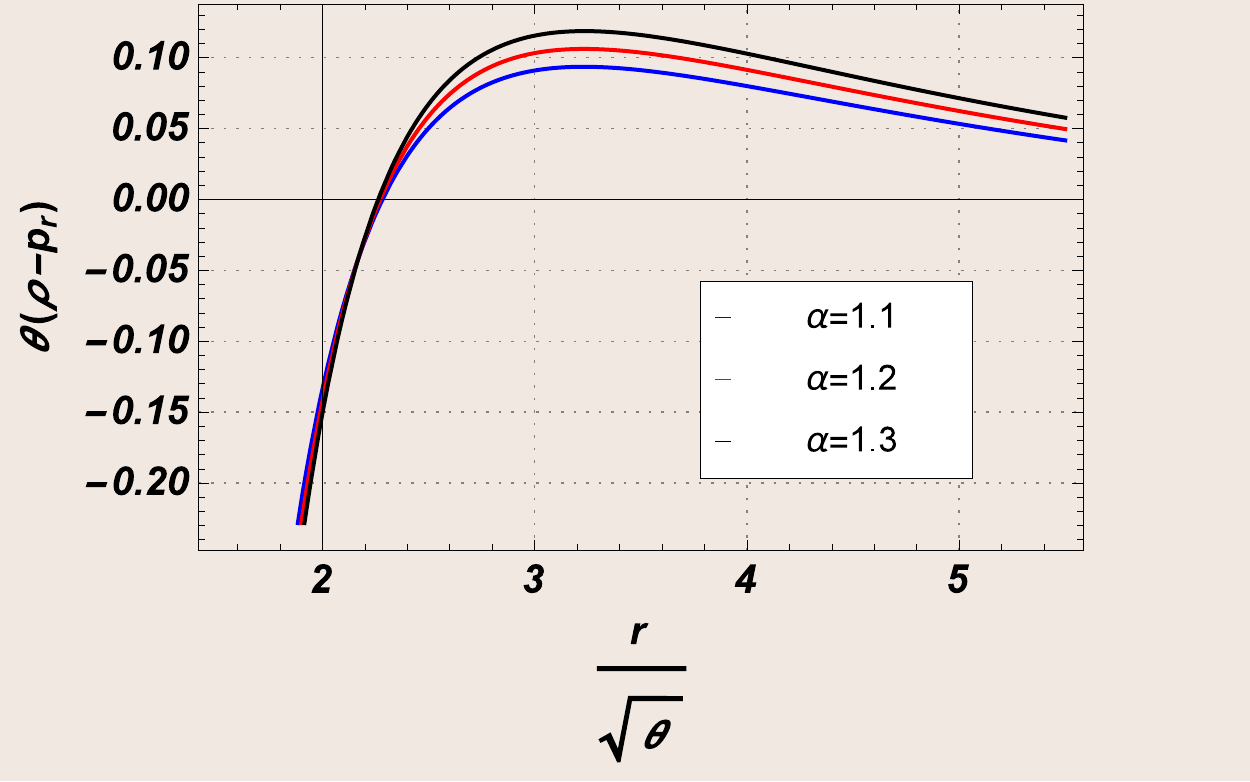, width=.48\linewidth,
height=2.5in}\epsfig{file=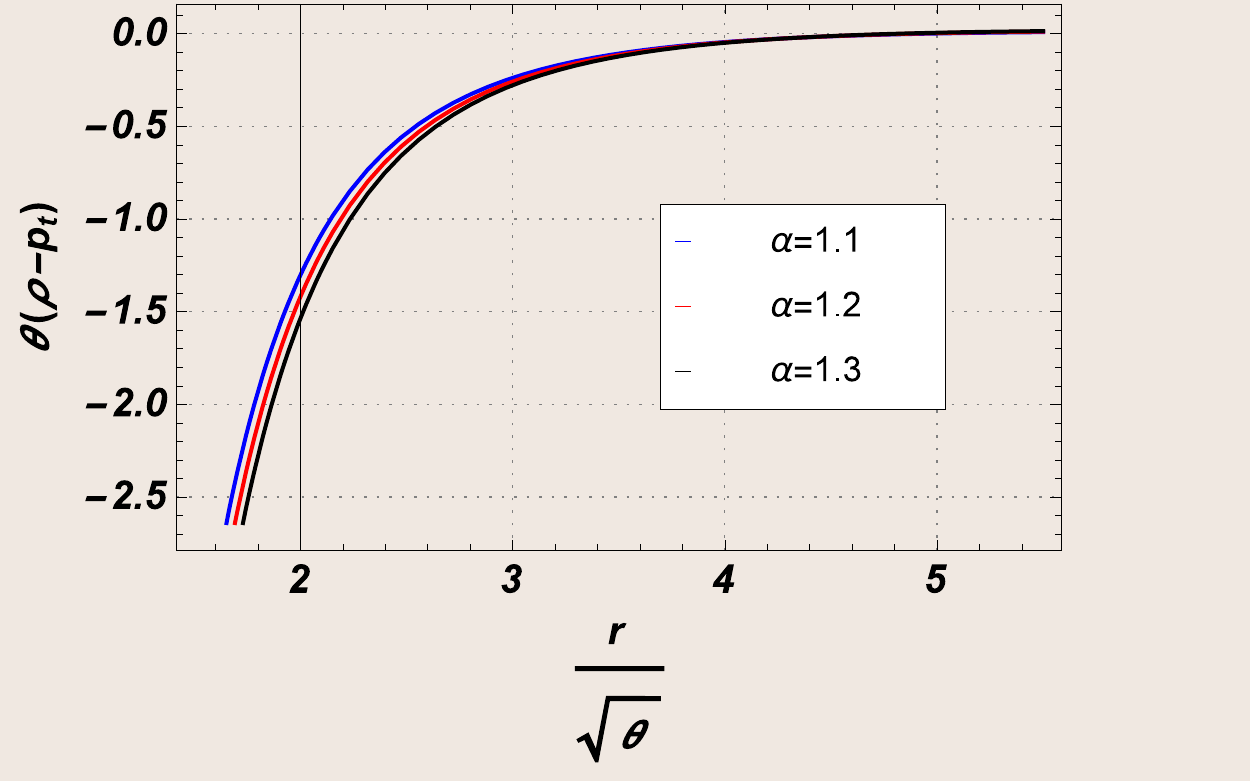, width=.48\linewidth,
height=2.5in}\caption{\label{Fig.5} Evolution of $\theta\left(\rho-p_r\right)$ and $\theta\left(\rho-p_t\right)$ with respect to $\frac{r}{\sqrt{\theta}}$ for different values of $\alpha$ with fixed parameters $M_1=0.2$, $\beta_1=0.07$, $K_2=1.2$ and $C_1=-5$ under Gaussian distribution.}
\end{figure}
\begin{figure}
\centering \epsfig{file=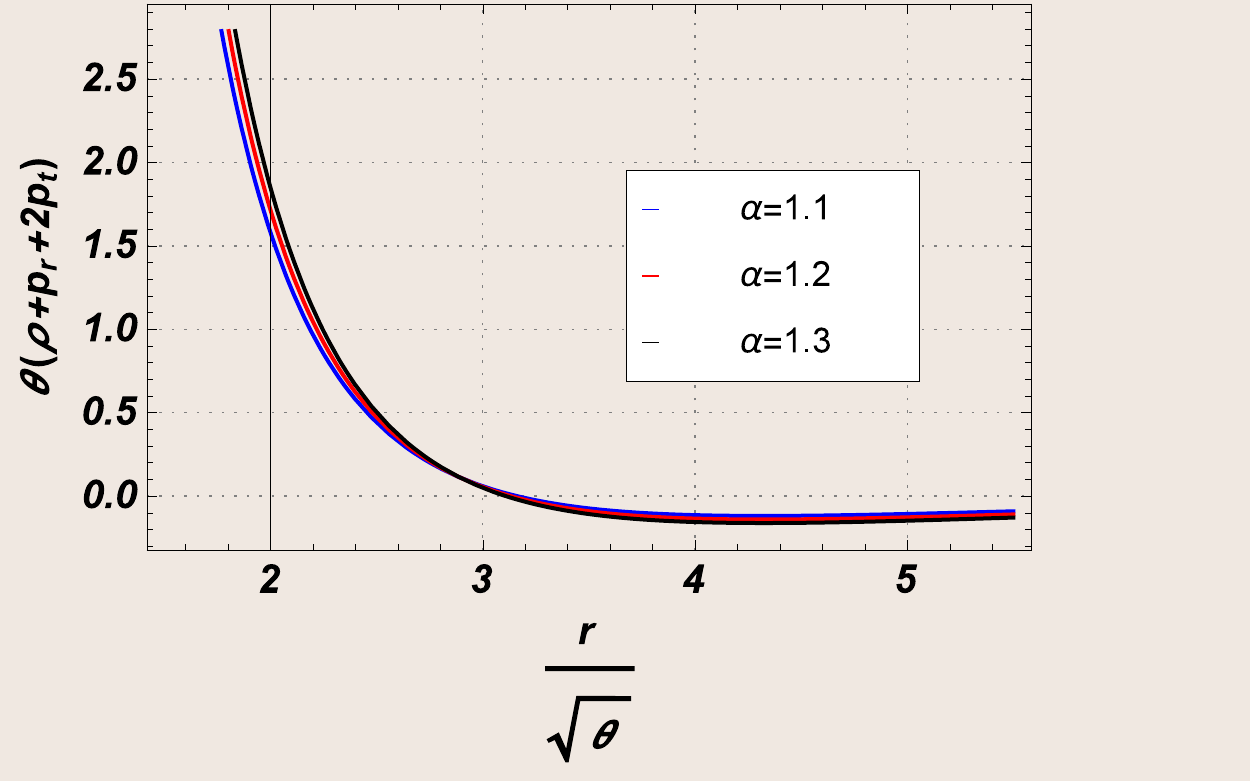, width=.48\linewidth,
height=2.5in}\caption{\label{Fig.6} Evolution of $\theta(\rho+p_r+2\,p_t)$ with respect to $\frac{r}{\sqrt{\theta}}$ for different values of $\alpha$ with fixed parameters $M_1=0.2$, $\beta_1=0.07$, $K_2=1.2$ and $C_1=-5$ under Gaussian distribution.}
\end{figure}
Now, we are going to discuss the behaviors of obtained shape function from Eq. \eqref{20} graphically. Here the shape function represent of the form $\frac{b}{\sqrt{\theta}}\left(\frac{r}{\sqrt{\theta}}\right)$ and this form of shape functions depends on $\alpha$, $M_1$, $\beta_1$, $K_2$ and $C_1$. Under this Gaussian sources framework we fixed some free parameters $M_1=0.2$, $\beta_1=0.07$, $K_2=1.2$ and $C_1=-5$ to analyse the results fruitfully. The left panel of Fig. \ref{Fig.1} shows the increasing behavior of $\frac{b}{\sqrt{\theta}}\left(\frac{r}{\sqrt{\theta}}\right)$ as $\left(\frac{r}{\sqrt{\theta}}\right)$ increases for different values of $\alpha$ and the right plot of Fig. \ref{Fig.1} indicates that $\frac{b^{'}}{\sqrt{\theta}}\left(\frac{r}{\sqrt{\theta}}\right)<1$ for $\left(\frac{r}{\sqrt{\theta}}\right)>\left(\frac{r_0}{\sqrt{\theta}}\right)$ i.e., flare-out condition satisfied. To check the asymptotically flatness condition we plot $\frac{\frac{b}{\sqrt{\theta}}}{\frac{r}{\sqrt{\theta}}}$ verses $\frac{r}{\sqrt{\theta}}$ which is presented in the left panel of Fig. \ref{Fig.2} and it is indicating that $\frac{\frac{b}{\sqrt{\theta}}}{\frac{r}{\sqrt{\theta}}}$ approaches to a small positive value as the radial coordinates $\frac{r}{\sqrt{\theta}}$  get larger values. Hence, it can be concluded that the asymptotically flatness behavior of the shape function cannot be achieved under this distribution. Also to locate the throat of wormhole, we imposed the condition $\frac{b}{\sqrt{\theta}}\left(\frac{r_0}{\sqrt{\theta}}\right)=\left(\frac{r_0}{\sqrt{\theta}}\right)$. The right panel of Fig. \ref{Fig.2} indicates that the wormhole's throat is located at $\frac{r}{\sqrt{\theta}}=\frac{r_0}{\sqrt{\theta}}$ where $\frac{b}{\sqrt{\theta}}\left(\frac{r}{\sqrt{\theta}}\right)-\left(\frac{r_0}{\sqrt{\theta}}\right)$ cuts the $\frac{r}{\sqrt{\theta}}$-axis. For $\alpha=1.1$, $\frac{b}{\sqrt{\theta}}\left(\frac{r}{\sqrt{\theta}}\right)-\left(\frac{r_0}{\sqrt{\theta}}\right)$ cuts the $\frac{r}{\sqrt{\theta}}$-axis  at $\left(\frac{r_0}{\sqrt{\theta}}\right)= 4.18$ (approximately). Whereas for $\alpha=1.2$ and $\alpha=1.3$, the area of throat radii located at $\left(\frac{r_0}{\sqrt{\theta}}\right)= 4.09$ and $\left(\frac{r_0}{\sqrt{\theta}}\right)= 4.02$ (approximately), respectively. One can notice that the position of the wormhole's throat decreasing with the increase of the parameter $\alpha$. In short, we can say that the shape function satisfy all the necessary condition for the WH under Gaussian distributions.\\
With the help of Eqs. \eqref{4a}, \eqref{21} and \eqref{22}, we have shown the behavior of energy conditions graphically in Figs. \ref{Fig.3}-\ref{Fig.6}. According to Fig. \ref{Fig.3}, $\,\,\theta(\rho)>0$ i.e. energy density is positive in the entire shape-time. In Figs. \ref{Fig.4} and \ref{Fig.5}, we have shown the behaviors of null energy condition NEC $\left(\theta(\rho+p_r),\,\,\,\theta(\rho+p_t)\right)$ and dominant energy condition DEC $\left(\theta(\rho-p_r),\,\,\,\theta(\rho-p_t)\right)$, respectively and the figures showing that both the energy conditions are violated. But the Fig. \ref{Fig.6} indicates that the strong energy condition SEC $\theta(\rho+p_r+2p_t)$) is satisfying. Violation of NEC hold the wormhole's throat open under this Gaussian distribution framework.

\subsection{Wormhole solutions under Lorentzian Distribution for $f(Q)$ gravity}
\label{sec4II}
In this subsection, we match the Lorentzian source of energy density \eqref{4b} with Eq. (\ref{15}), we get the following differential equation
\begin{equation}
\frac{\sqrt{\theta } M}{\pi ^2 \left(\theta +r^2\right)^2}=-\frac{\beta}{2}+\frac{\alpha}{r^2}-\frac{\alpha\,F(r)}{K_2^2\,r^2}\left(2\,r\,F^{'}(r)+F(r)\right),\label{23}
\end{equation}
After solving the above differential equation with the scope of Eq. \eqref{13}, we get the following shape function
\begin{equation}
\frac{b}{\sqrt{\theta }}\left(\frac{r}{\sqrt{\theta }}\right)=\frac{C_{2}}{K^{2}_{2}}-\frac{K_2^2}{6\alpha}\left(\frac{r}{\sqrt{\theta }}\right)\left(\frac{3 M_1}{\pi ^2 \left(\left(\frac{r}{\sqrt{\theta }}\right)^2+1\right)}+\beta_{1} \left(\frac{r}{\sqrt{\theta }}\right)^2-5\alpha\right)-\frac{K^{2} _{2} M_1 \cot ^{-1}\left(\frac{r}{\sqrt{\theta }}\right)}{2\,\pi ^2 \alpha },\label{24}
\end{equation}
where $C_{2}=\sqrt{\theta }C_{00}$ with $C_{00}$ is a constant of integration for Eq. (\ref{23}).  By using the Eq. (\ref{20}) in Eq. (\ref{16}) and Eq. (\ref{17}), we get a following expressions for pressure components
\begin{equation}
\label{25}
p_r=\frac{1}{\theta}\left(\frac{\beta_1}{3}+\frac{1}{2\,\left(\frac{r}{\sqrt{\theta}}\right)^3}\left(-\frac{10 \alpha  C_{2}}{K^{4} _{2}}+\frac{5 M_1 r}{\sqrt{\theta } \left(\pi ^2 \left(\left(\frac{r}{\sqrt{\theta }}\right)^2+1\right)\right)}-\frac{8 \alpha  r}{\sqrt{\theta }}+\frac{5 M_1 \cot ^{-1}\left(\frac{r}{\sqrt{\theta }}\right)}{\pi^2}\right)\right),
\end{equation}
\begin{eqnarray}
p_t&=&\frac{1}{\theta}\frac{1}{12 \pi ^2 K_{2} ^{4} \left(\frac{r}{\sqrt{\theta}}\right)^3 \left(\left(\frac{r}{\sqrt{\theta}}\right)^2+1\right)^2}\bigg(-30 \pi ^2 \alpha  C_{2} \left(\left(\frac{r}{\sqrt{\theta }}\right)^2+1\right)^2+K^{4} _{2}\frac{r}{\sqrt{\theta }}\left(-3 M_1 \left(\left(\frac{r}{\sqrt{\theta }}\right)^2-5\right)\right.\nonumber\\&+&\left.8 \pi ^2 \left(\left(\frac{r}{\sqrt{\theta }}\right)^2+1\right)^2 \left(\beta_{1} \left(\frac{r}{\sqrt{\theta }}\right)^2-6 \alpha \right)\right)+15 K^{4} _{2} M_1 \left(\left(\frac{r}{\sqrt{\theta }}\right)^2+1\right)^2 \cot ^{-1}\left(\frac{r}{\sqrt{\theta }}\right)\bigg)\label{26}.
\end{eqnarray}
\begin{figure}
\centering \epsfig{file=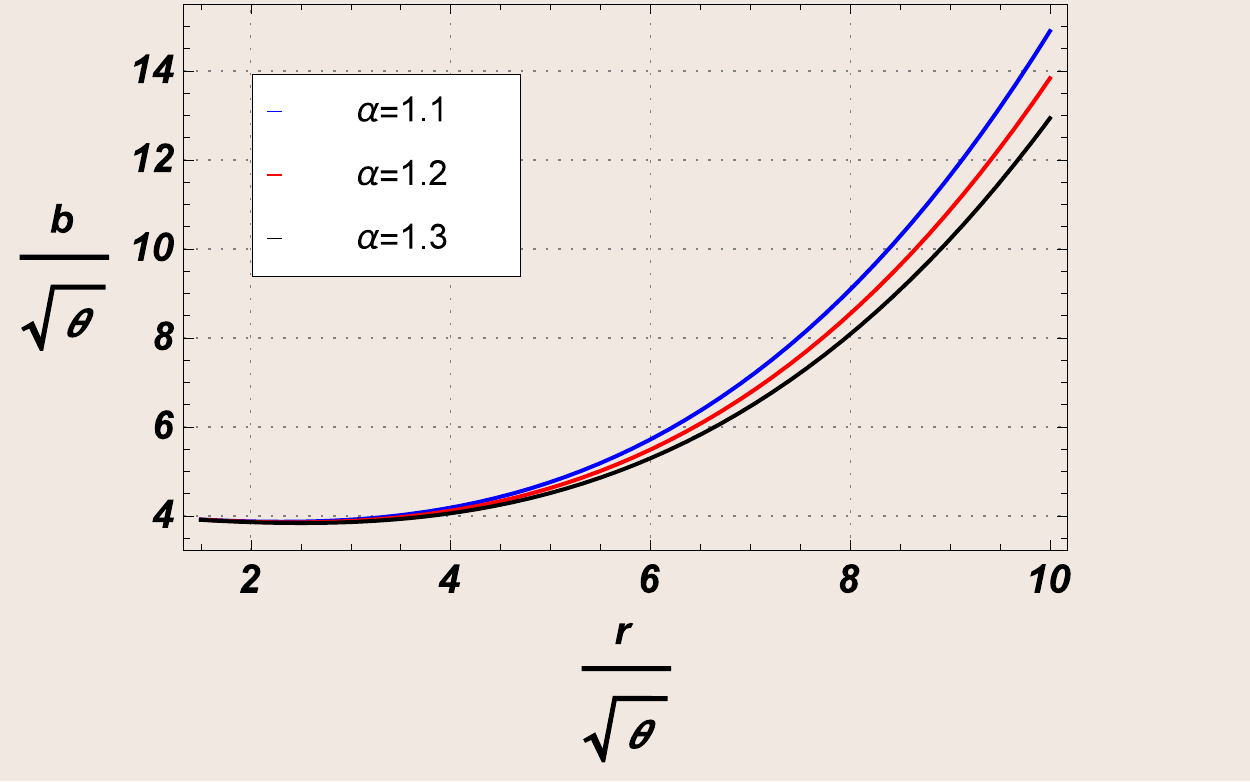, width=.48\linewidth,
height=2.5in}\epsfig{file=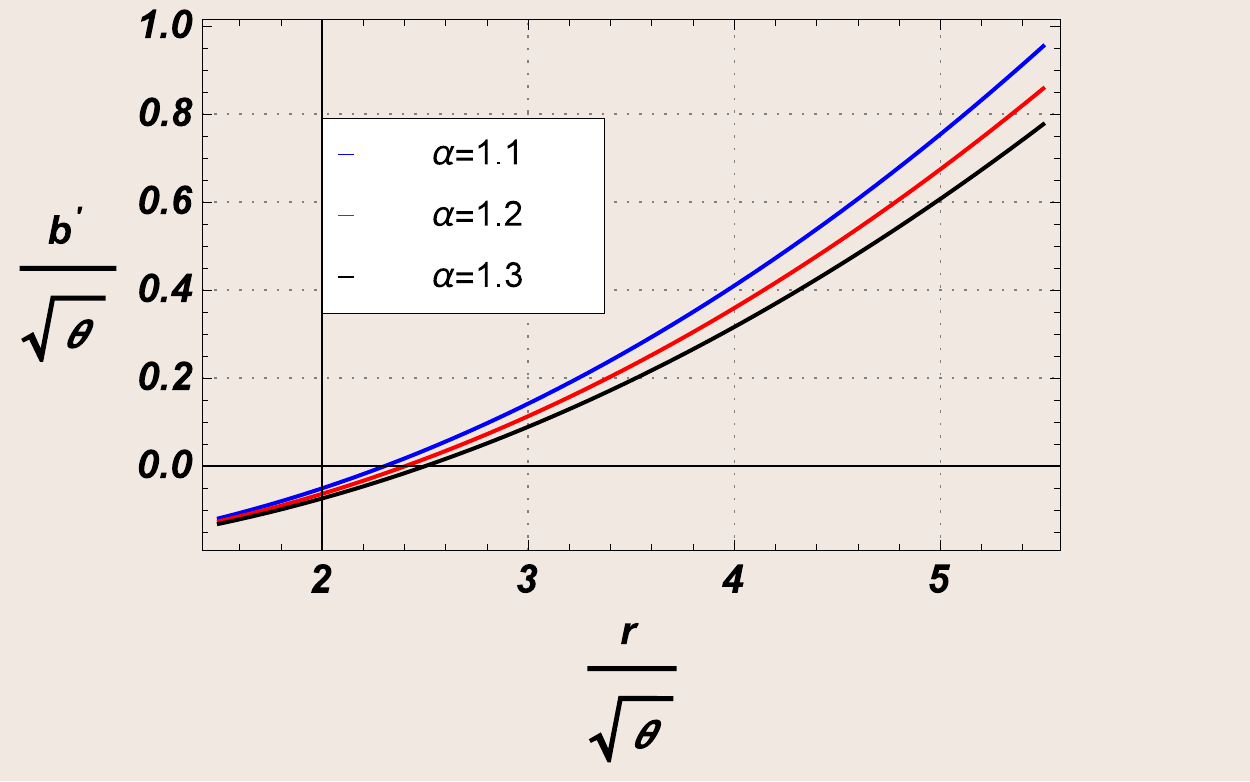, width=.48\linewidth,
height=2.5in}\caption{\label{Fig.7} Shows the behavior of $\frac{b}{\sqrt{\theta}}\left(\frac{r}{\sqrt{\theta}}\right)$ and $\frac{b^{'}}{\sqrt{\theta}}\left(\frac{r}{\sqrt{\theta}}\right)$ with respect to $\frac{r}{\sqrt{\theta}}$ for different values of $\alpha$ with fixed parameters $M_1=0.2$, $\beta_1=0.07$, $K_2=1.2$ and $C_2=-5$ under Lorentzian distribution.}
\end{figure}
\begin{figure}
\centering \epsfig{file=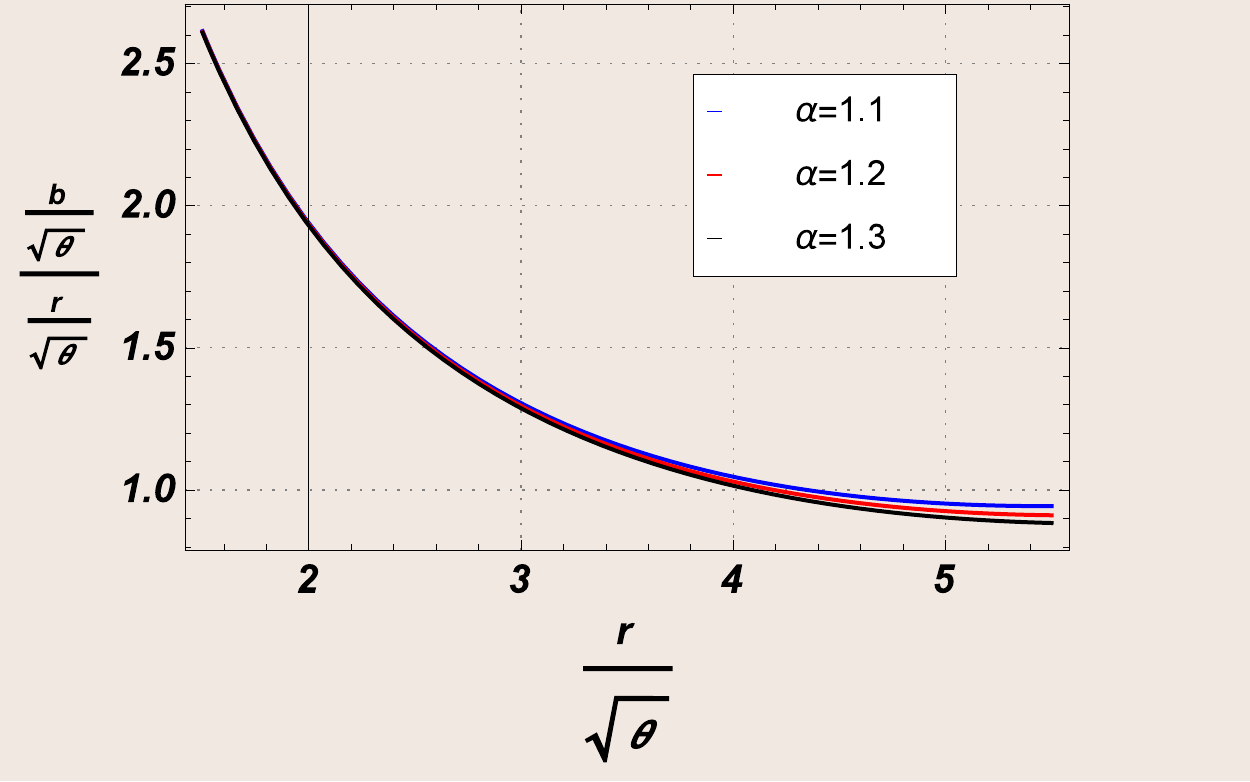, width=.48\linewidth,
height=2.5in}\epsfig{file=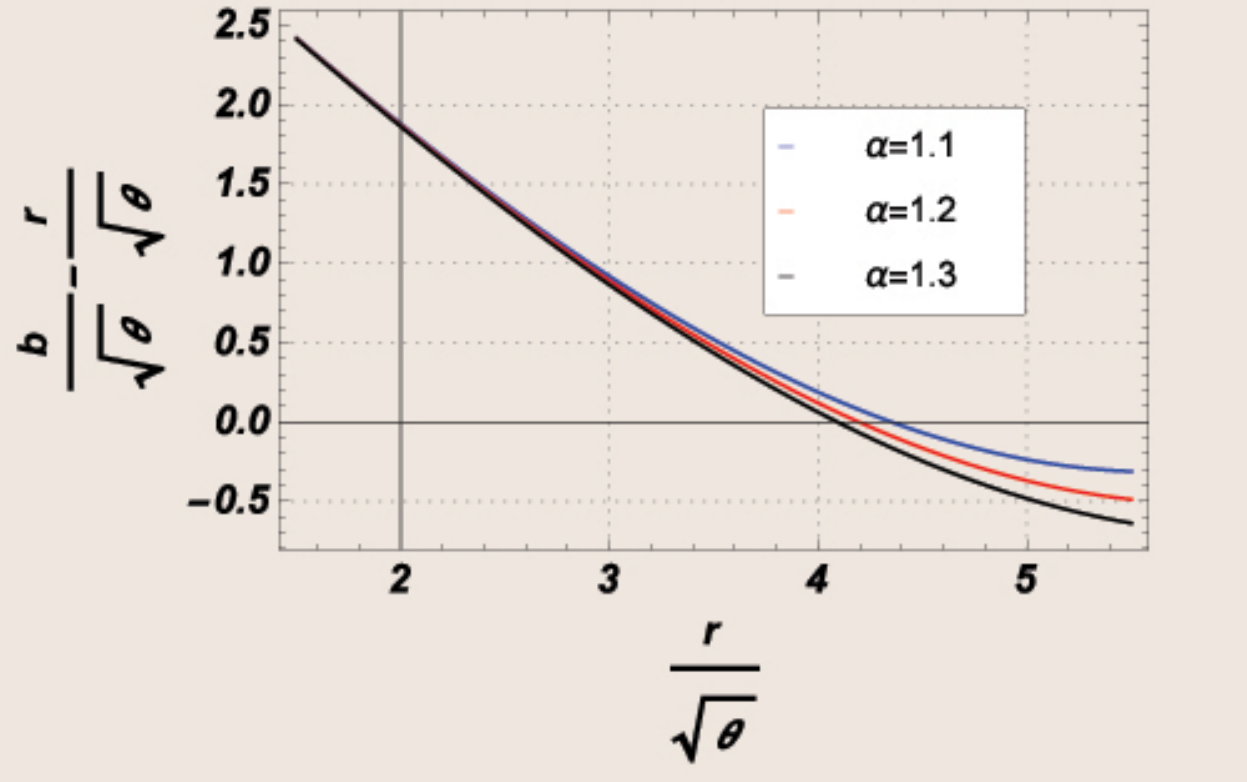, width=.48\linewidth,
height=2.5in}\caption{\label{Fig.8} Shows the behavior of $\frac{\frac{b}{\sqrt{\theta}}\left(\frac{r}{\sqrt{\theta}}\right)}{\frac{r}{\sqrt{\theta}}}$ and $\frac{b}{\sqrt{\theta}}\left(\frac{r}{\sqrt{\theta}}\right)-\left(\frac{r_0}{\sqrt{\theta}}\right)$ with respect to $\frac{r}{\sqrt{\theta}}$ for different values of $\alpha$ with fixed parameters $M_1=0.2$, $\beta_1=0.07$, $K_2=1.2$ and $C_2=-5$ under Lorentzian distribution.}
\end{figure}
\begin{figure}
\centering \epsfig{file=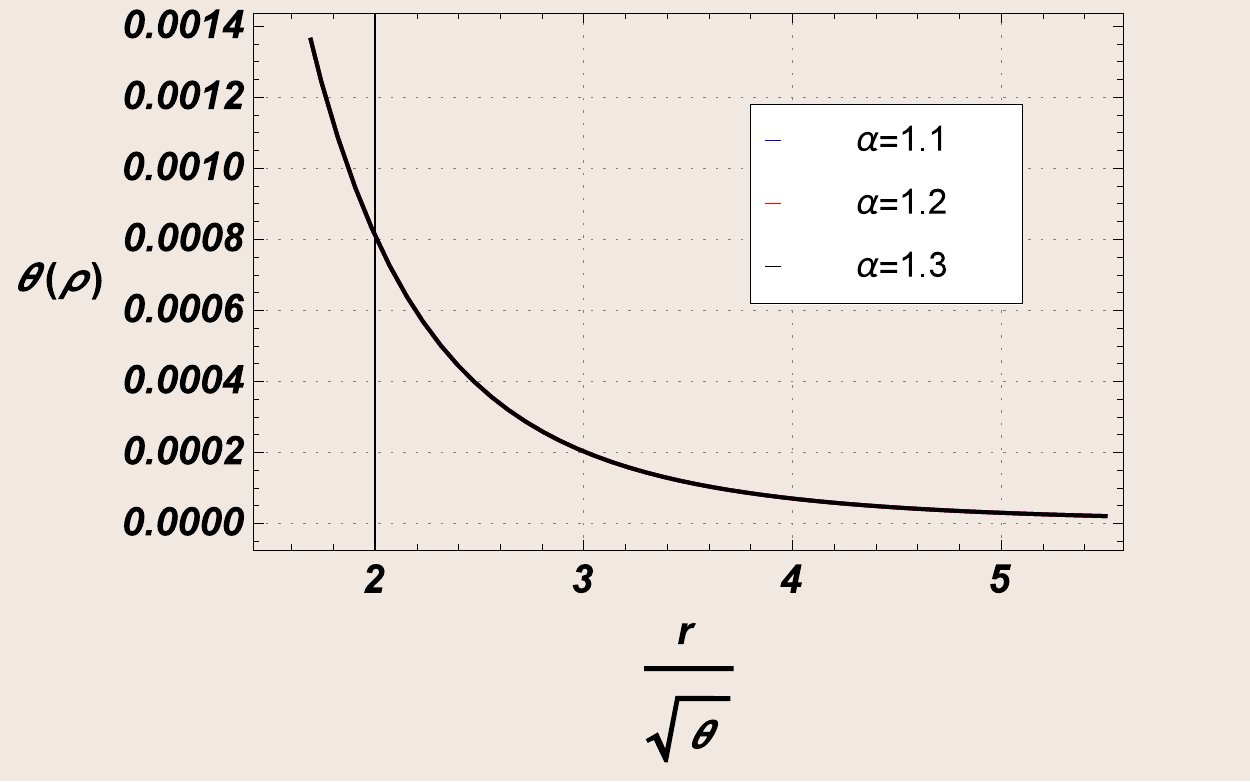, width=.48\linewidth,
height=2.5in}\caption{\label{Fig.9} Evolution of $\theta(\rho)$ with respect to $\frac{r}{\sqrt{\theta}}$ for different values of $\alpha$ with fixed parameters $M_1=0.2$, $\beta_1=0.07$, $K_2=1.2$ and $C_2=-5$ under Lorentzian distribution.}
\end{figure}
\begin{figure}
\centering \epsfig{file=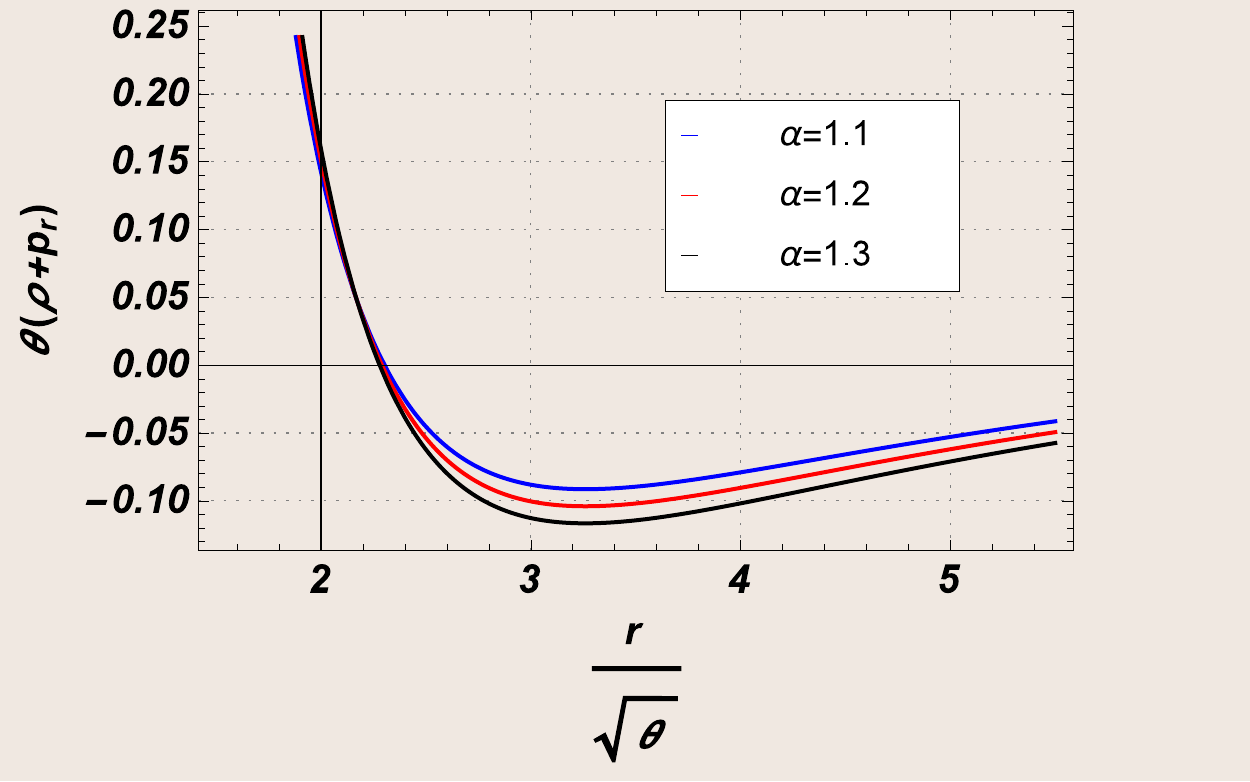, width=.48\linewidth,
height=2.5in}\epsfig{file=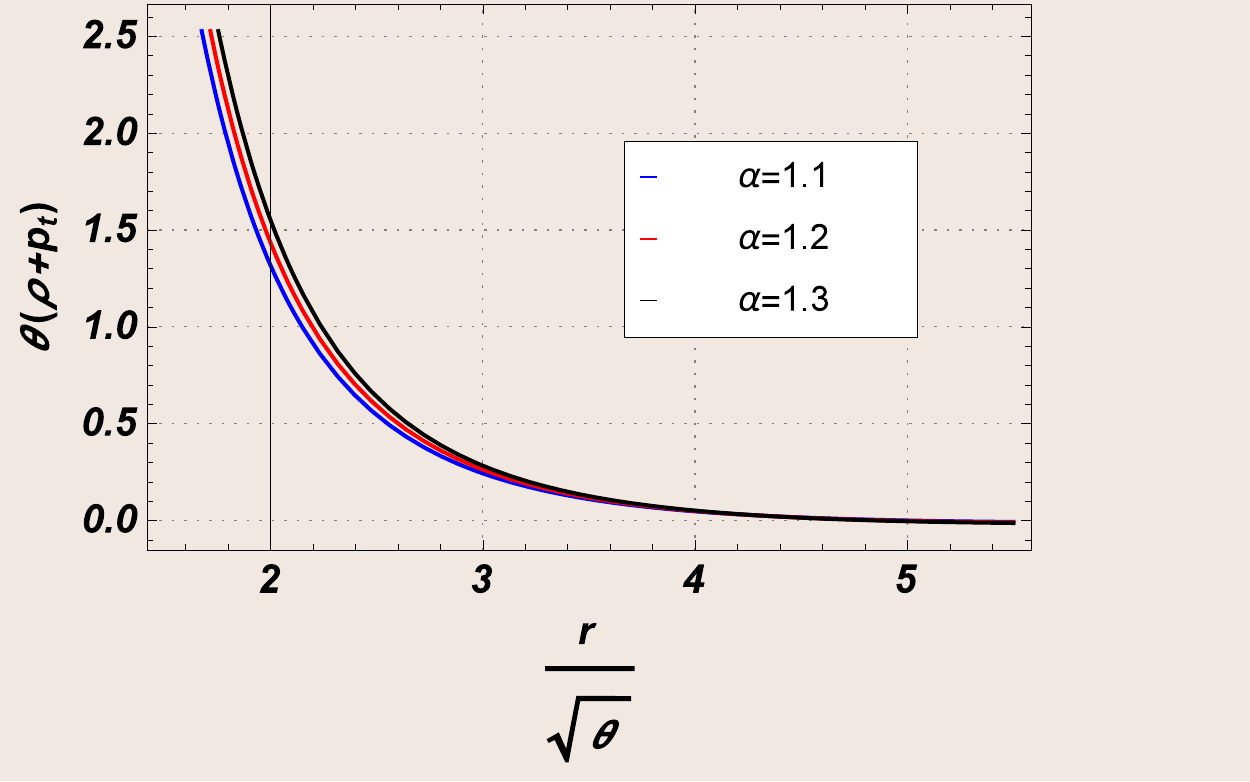, width=.48\linewidth,
height=2.5in}\caption{\label{Fig.10} Evolution of $\theta\left(\rho+p_r\right)$ and $\theta\left(\rho+p_t\right)$ with respect to $\frac{r}{\sqrt{\theta}}$ for different values of $\alpha$ with fixed parameters $M_1=0.2$, $\beta_1=0.07$, $K_2=1.2$ and $C_2=-5$ under Lorentzian distribution.}
\end{figure}
\begin{figure}
\centering \epsfig{file=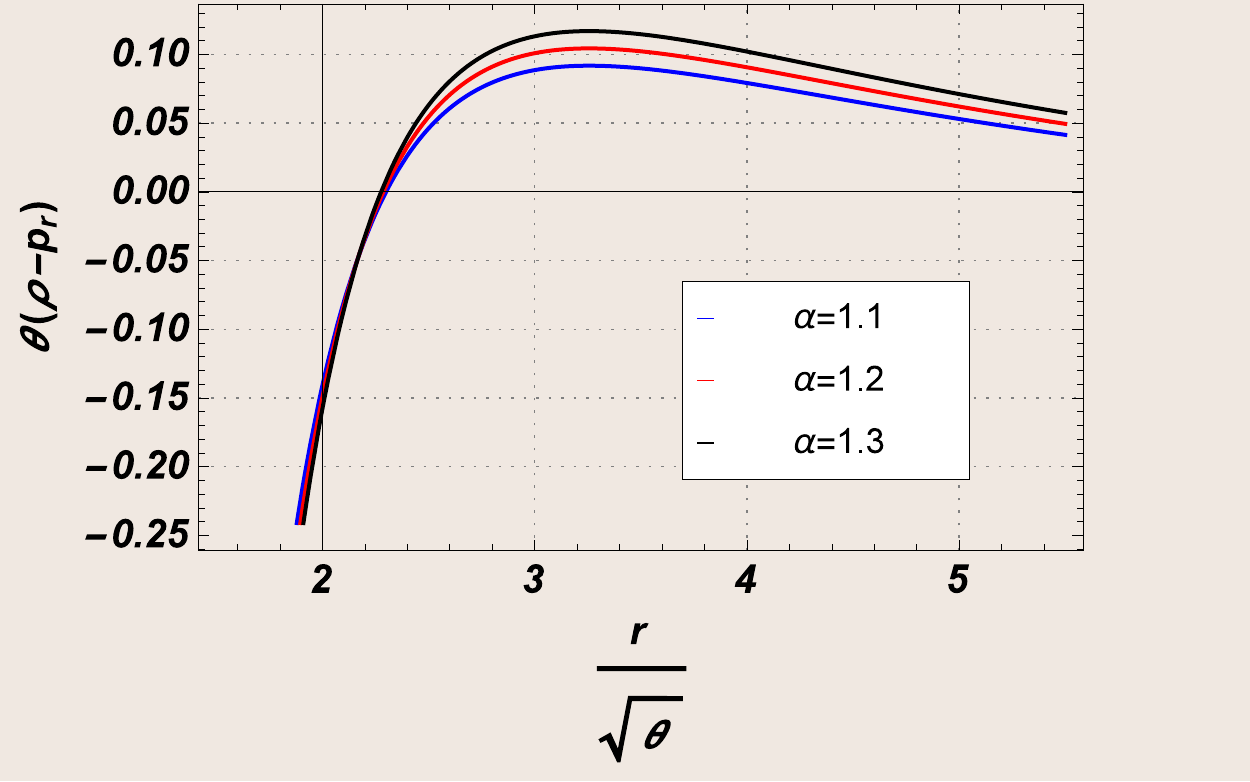, width=.48\linewidth,
height=2.5in}\epsfig{file=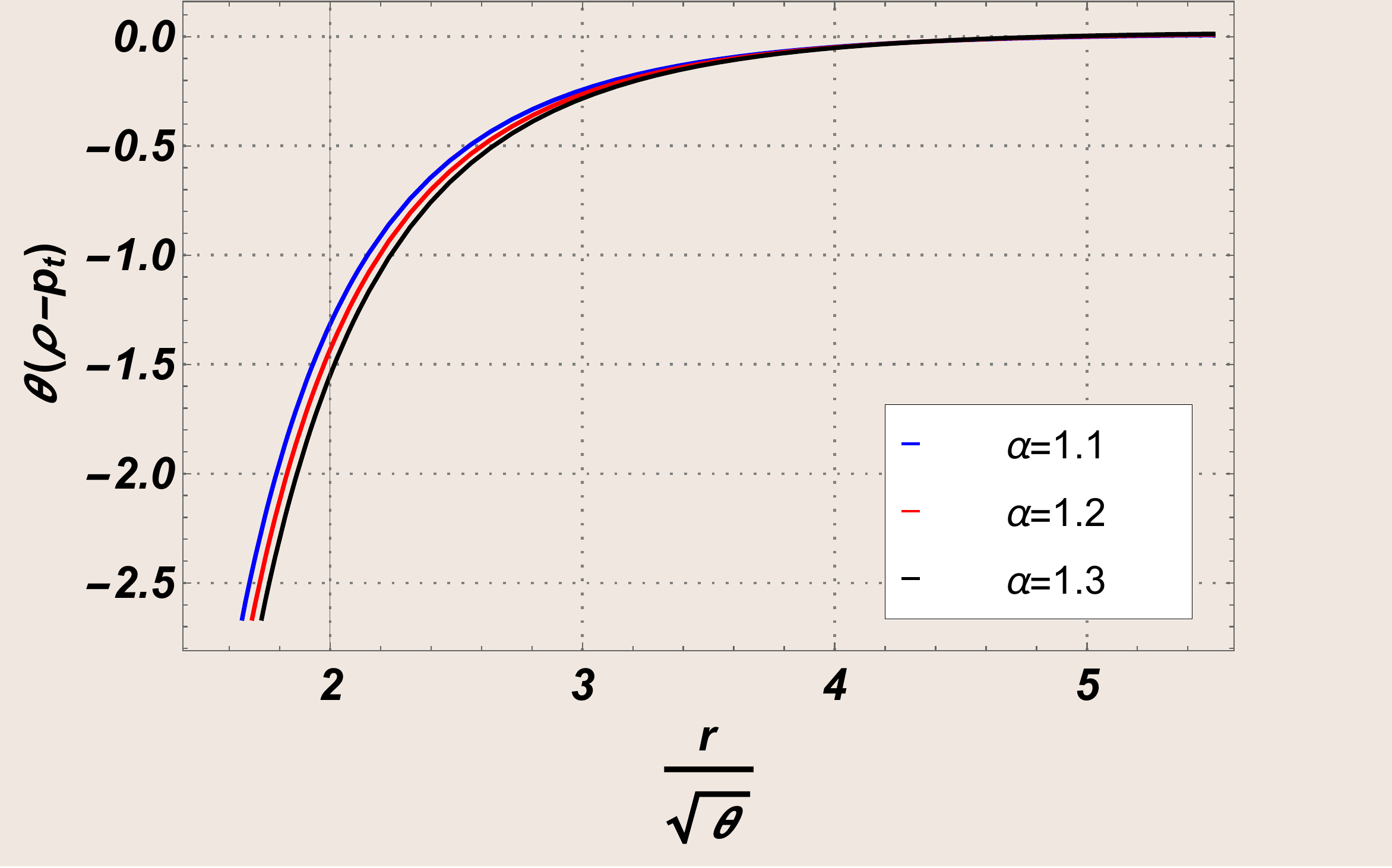, width=.48\linewidth,
height=2.5in}\caption{\label{Fig.11} Evolution of $\theta\left(\rho-p_r\right)$ and $\theta\left(\rho-p_t\right)$ with respect to $\frac{r}{\sqrt{\theta}}$ for different values of $\alpha$ with fixed parameters $M_1=0.2$, $\beta_1=0.07$, $K_2=1.2$ and $C_2=-5$ under Lorentzian distribution.}
\end{figure}
\begin{figure}
\centering \epsfig{file=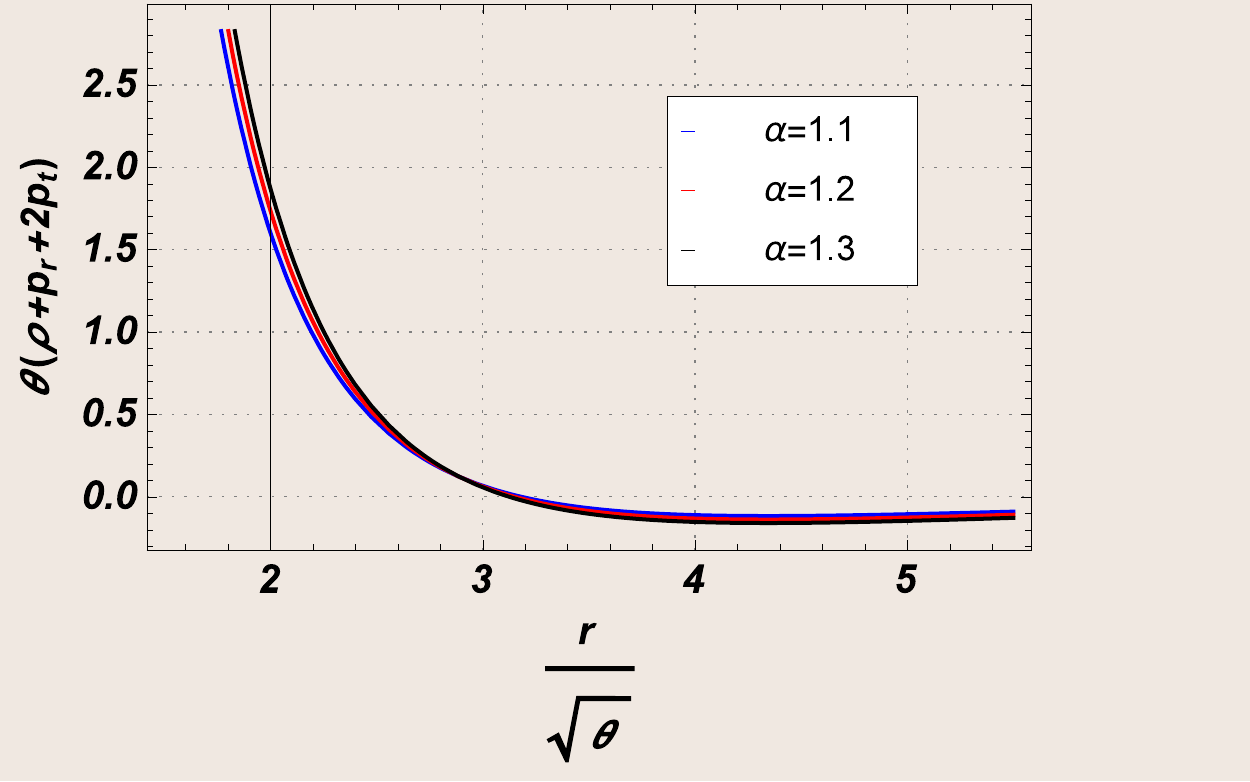, width=.48\linewidth,
height=2.5in}\caption{\label{Fig.12} Evolution of $\theta(\rho+p_r+2\,p_t)$ with respect to $\frac{r}{\sqrt{\theta}}$ for different values of $\alpha$ with fixed parameters $M_1=0.2$, $\beta_1=0.07$, $K_2=1.2$ and $C_2=-5$ under Lorentzian distribution.}
\end{figure}
In Figs. \ref{Fig.7} and \ref{Fig.8}, we have discussed the behaviors of obtained shape function from Eq. \eqref{24} graphically. In this case, for a best fit configuration we set the free parameters similar to subsection \ref{sec4I}. The left panel of Fig. \ref{Fig.7} indicates that the shape function $\frac{b}{\sqrt{\theta}}\left(\frac{r}{\sqrt{\theta}}\right)$ increases as $\left(\frac{r}{\sqrt{\theta}}\right)$ increases and the right plot of Fig. \ref{Fig.7} shows that flare-out condition satisfied i.e., $\frac{b^{'}}{\sqrt{\theta}}\left(\frac{r}{\sqrt{\theta}}\right)<1$ for $\left(\frac{r}{\sqrt{\theta}}\right)>\left(\frac{r_0}{\sqrt{\theta}}\right)$ for different values of $\alpha$. We have plotted the graph of $\frac{\frac{b}{\sqrt{\theta}}}{\frac{r}{\sqrt{\theta}}}$ with respect to $\frac{r}{\sqrt{\theta}}$ in the left panel of Fig. \ref{Fig.8}, to check the asymptotically flatness condition and the graph indicating that $\frac{\frac{b}{\sqrt{\theta}}}{\frac{r}{\sqrt{\theta}}}$ tends to a small positive value as $\frac{r}{\sqrt{\theta}}$ increases which means that flatness condition is not satisfied under the lorentzian distribution also. Moreover, to find the wormhole throat, we plot the graph of $\frac{b}{\sqrt{\theta}}\left(\frac{r}{\sqrt{\theta}}\right)-\left(\frac{r_0}{\sqrt{\theta}}\right)$ verses $\frac{r}{\sqrt{\theta}}$ in the right panel of Fig. \ref{Fig.8} which indicates that the wormhole throat is located at $\frac{r}{\sqrt{\theta}}=\frac{r_0}{\sqrt{\theta}}$ where $\frac{b}{\sqrt{\theta}}\left(\frac{r}{\sqrt{\theta}}\right)-\left(\frac{r_0}{\sqrt{\theta}}\right)$ cuts the $\frac{r}{\sqrt{\theta}}$-axis. For $\alpha=1.1$, $\frac{b}{\sqrt{\theta}}\left(\frac{r}{\sqrt{\theta}}\right)-\left(\frac{r_0}{\sqrt{\theta}}\right)$ crosses the horizontal axis $\frac{r}{\sqrt{\theta}}$  at $\left(\frac{r_0}{\sqrt{\theta}}\right)= 4.20$ (approximately). Similarly, for $\alpha=1.2$ and $\alpha=1.3$, the area of throat radii located at $\left(\frac{r_0}{\sqrt{\theta}}\right)= 4.10$ and $4.08$ (approximately), respectively. In this case also we found that the position of the wormhole throat decreasing with the increase of the parameter $\alpha$. Hence, under this Lorentzian distribution, shape function satisfy all the basic criteria for a traversable wormhole.\\
We are now using Eqs. \eqref{4b}, \eqref{25} and \eqref{26}, to show the behavior of energy conditions graphically in Figs. \ref{Fig.9}-\ref{Fig.12}. From Fig. \ref{Fig.9}, we can say that the energy density is positive throughout the shape-time. In Figs. \ref{Fig.10}, the profile of NEC is depicted where one can observe that $\theta(\rho+p_r)<0$, i.e., NEC is violated. Violation of NEC may confirm the exotic matters at wormhole's throat. Moreover, the behaviors of DEC  and SEC have been shown in Figs. \ref{Fig.11} and \ref{Fig.12}, respectively. In this case, we have found that DEC is violated while the SEC is obeyed.
\section{Equilibrium Condition}
\label{sec5}
In order to find the equilibrium configuration for the wormhole geometry in the background of noncommutative Gaussian and Lorentzian distributions, we shall use the generalized Tolman-Oppenheimer-Volkoff (TOV) equation. The generalized TOV equation is provided as \cite{Khufitting/2014,Ray/2014}
\begin{equation}\label{27}
-\frac{dp_{r}}{dr}-\frac{\epsilon^{'}(r)}{2}(\rho+p_{r})+\frac{2}{r}(p_{t}-p_{r})=0,
\end{equation}
where $\epsilon(r)=2\Psi(r)$. The forces namely, hydrostatic $(\mathcal{F}_{\mathrm{h}})$, the gravitational ($\mathcal{F}_{\mathrm{g}})$ and anisotropic force $(\mathcal{F}_{\mathrm{a}})$ are represented by following expressions
\begin{equation}\label{28}
\mathcal{F}_{\mathrm{h}}=-\frac{dp_{r}}{dr},\;\;\;\;\;\;\;\;\mathcal{F}_{\mathrm{a}}=\frac{2}{r}(p_{t}-p_{r}), \;\;\;\;\;\;\;\;\mathcal{F}_{\mathrm{g}}=-\frac{\epsilon^{'}}{2}(\rho+p_{r}),
\end{equation}
and thus Eq. \eqref{27} takes the form given by
\begin{equation*}
\mathcal{F}_{\mathrm{a}}+\mathcal{F}_{\mathrm{g}}+\mathcal{F}_{\mathrm{h}}=0.
\end{equation*}

\begin{figure}
\centering \epsfig{file=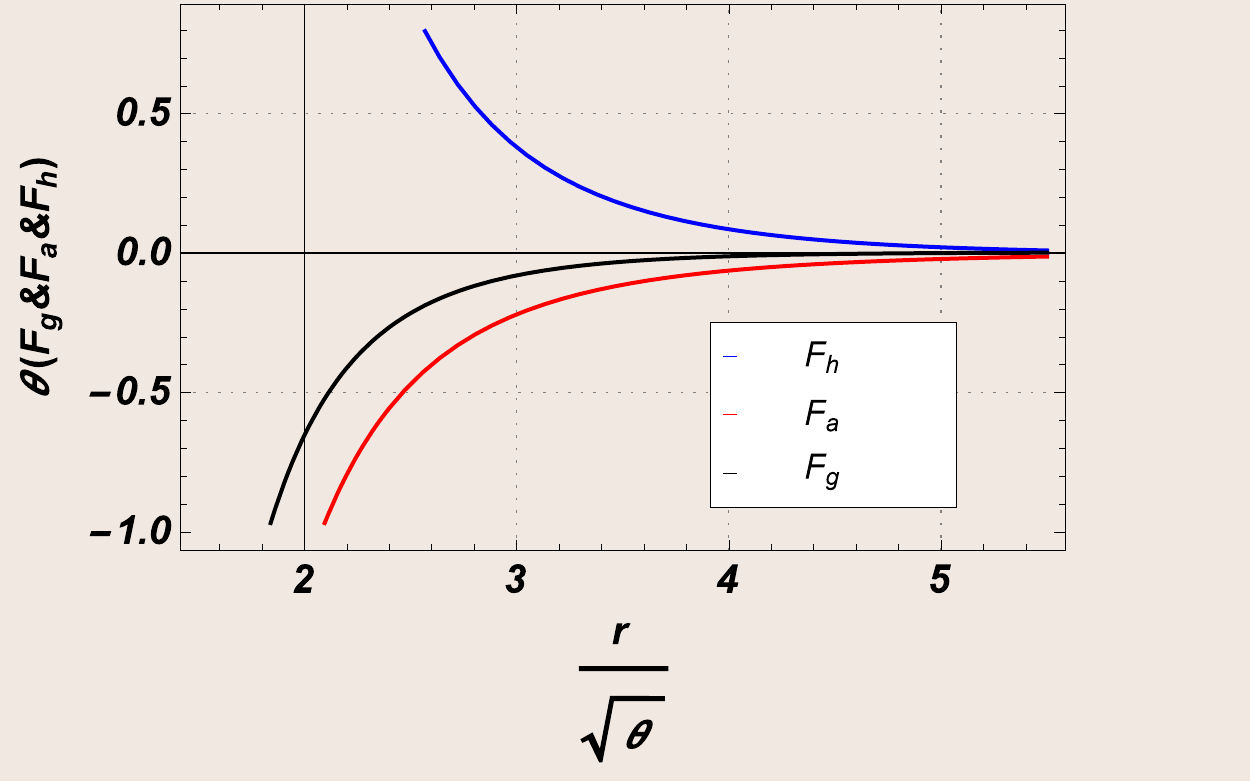, width=.32\linewidth,
height=2.5in}\epsfig{file=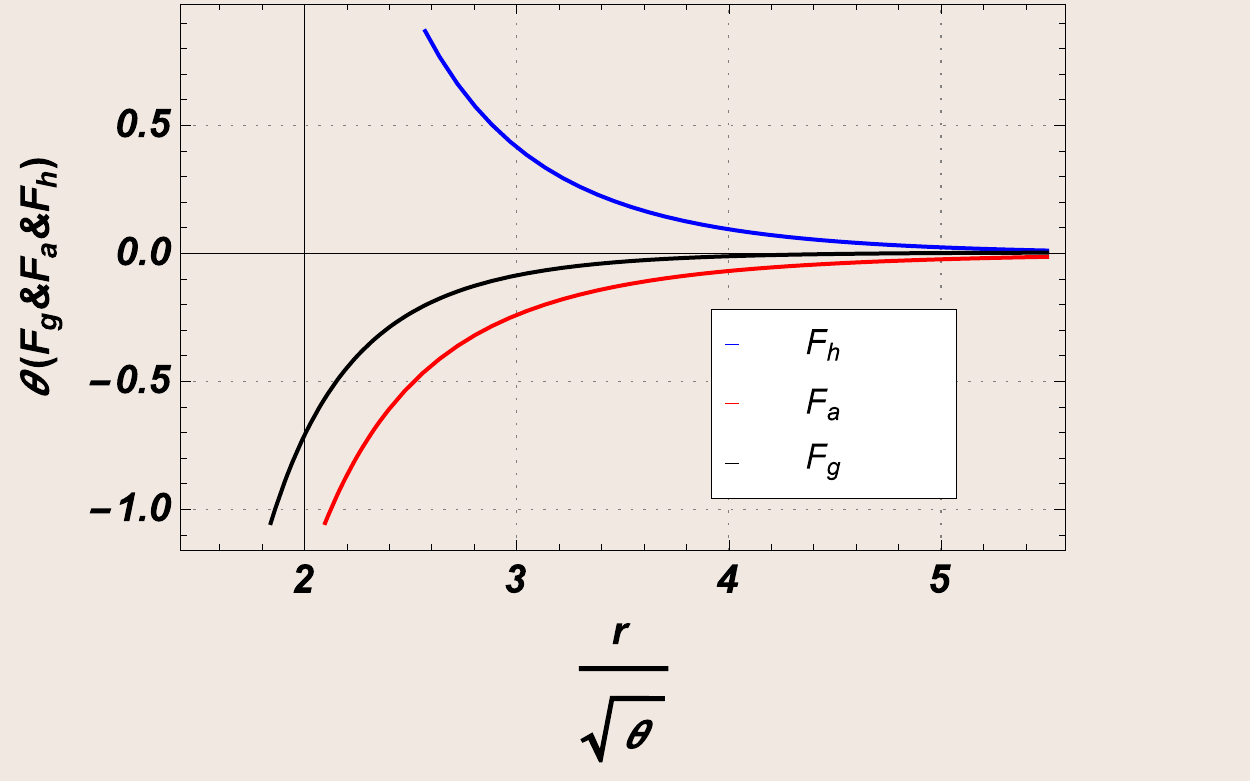, width=.32\linewidth,
height=2.5in}\epsfig{file=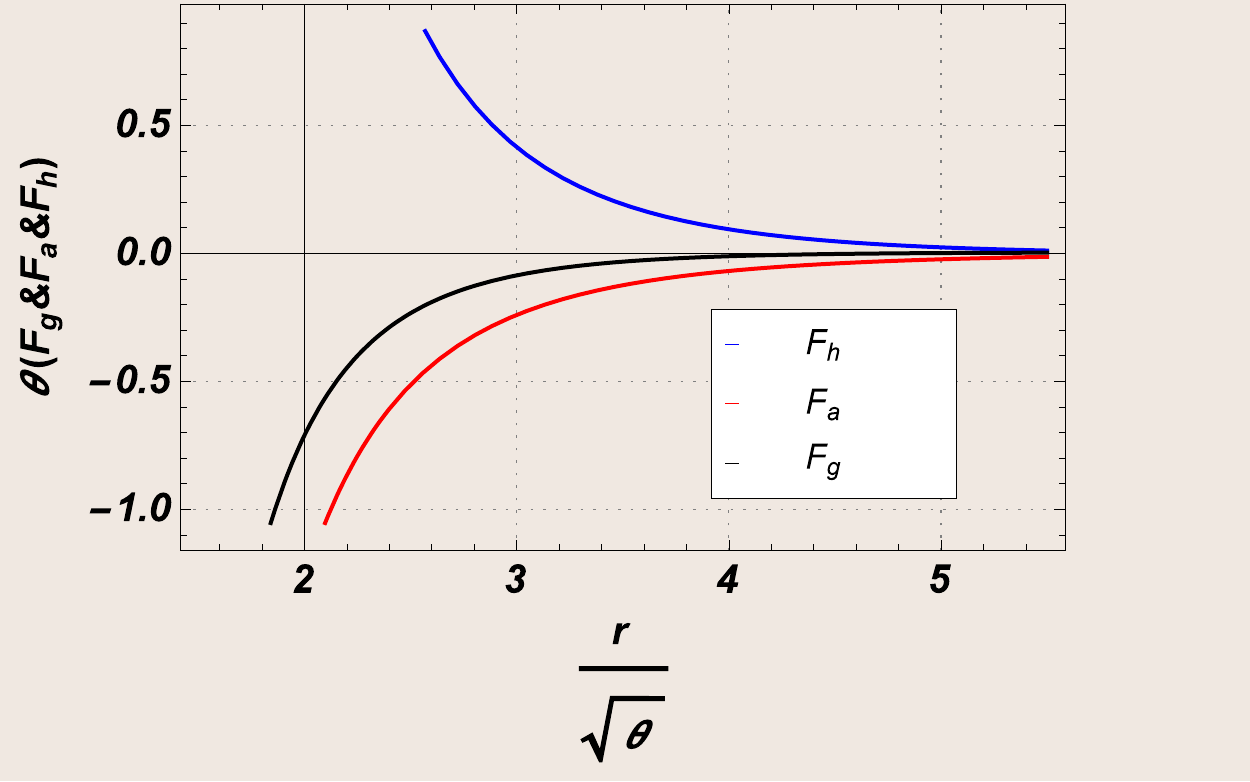, width=.32\linewidth,
height=2.5in}\caption{\label{Fig.13} This profile shows the graphical representation of $F_h$, $F_a$ and $F_g$ with respect to $\frac{r}{\sqrt{\theta}}$ for $\alpha=1.1\,\,(left)$, $\alpha=1.2\,\,(middle)$ and $\alpha=1.3\,\,(right)$ with fixed parameters $M_1=0.2$, $\beta_1=0.07$, $K_1=K_2=1.2$ and $C_1=-5$ under Gaussian distribution.}
\end{figure}

\begin{figure}
\centering \epsfig{file=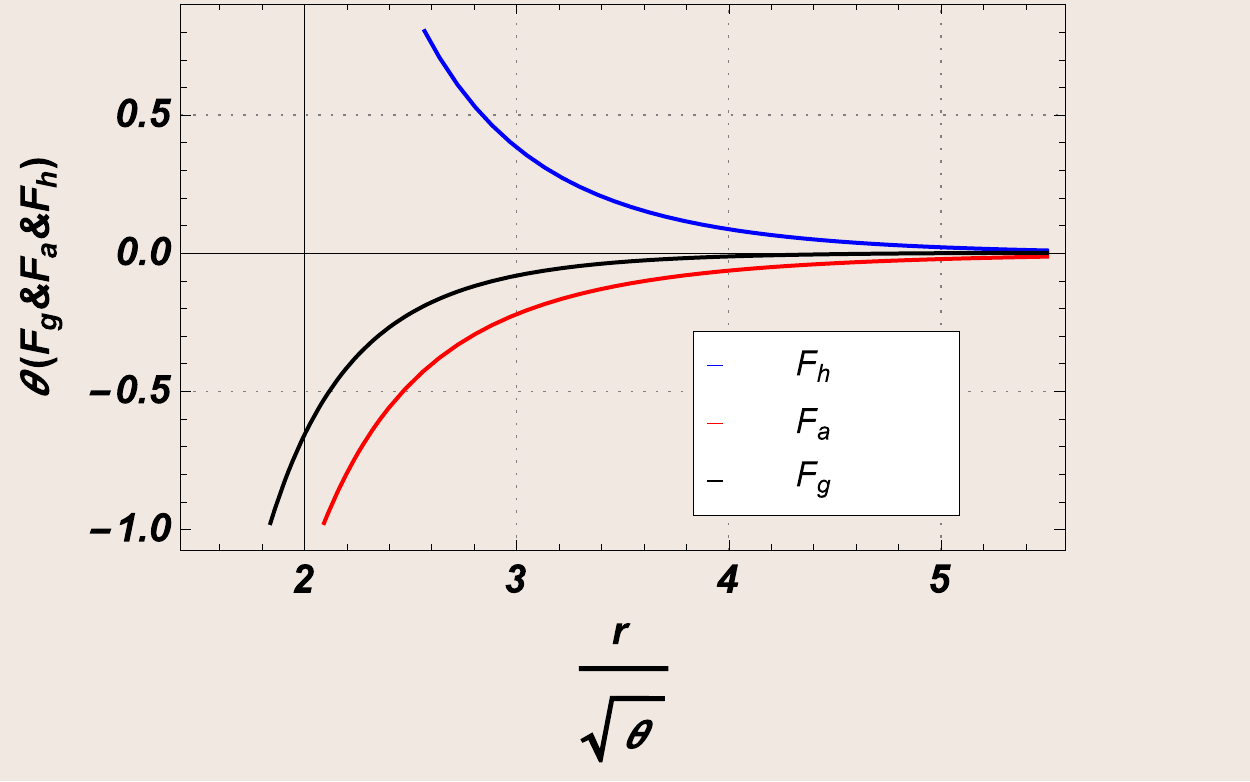, width=.32\linewidth,
height=2.5in}\epsfig{file=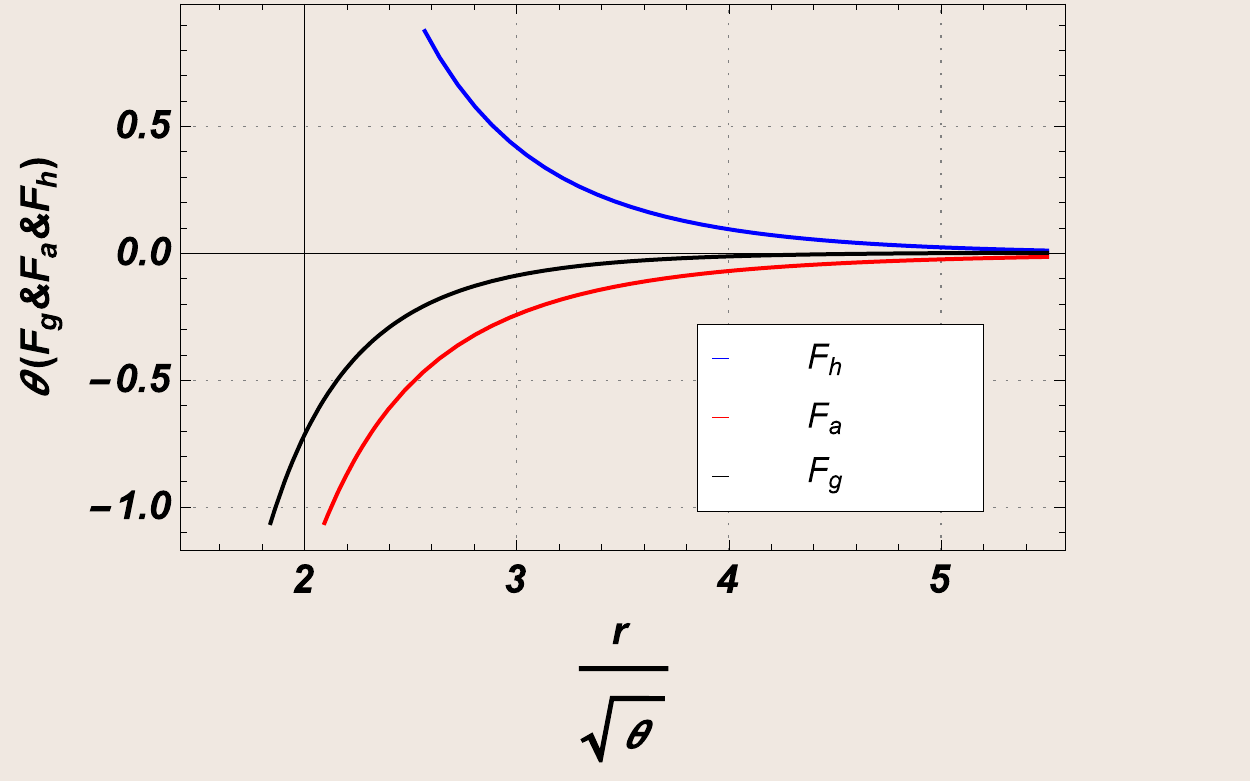, width=.32\linewidth,
height=2.5in}\epsfig{file=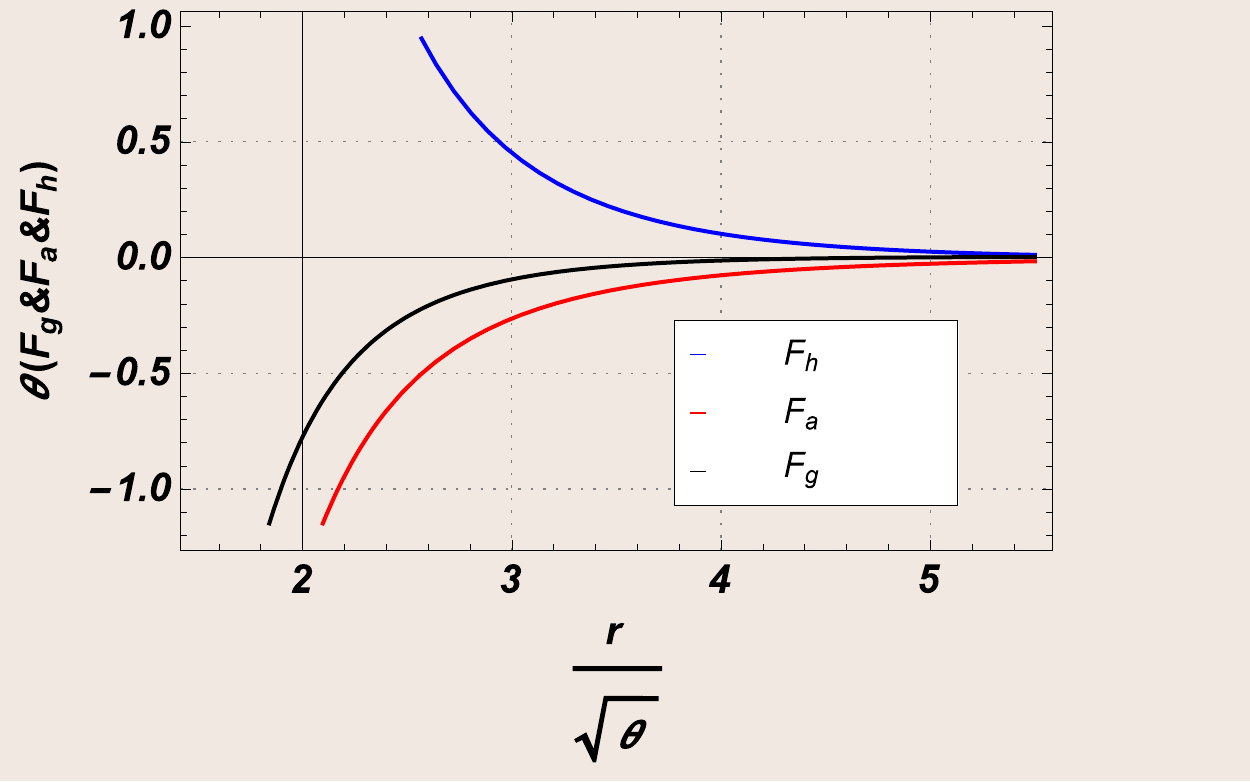, width=.32\linewidth,
height=2.5in}\caption{\label{Fig.14} This profile shows the graphical representation of $F_h$, $F_a$ and $F_g$ with respect to $\frac{r}{\sqrt{\theta}}$ for $\alpha=1.1\,\,(left)$, $\alpha=1.2\,\,(middle)$ and $\alpha=1.3\,\,(right)$ with fixed parameters $M_1=0.2$, $\beta_1=0.07$, $K_1=K_2=1.2$ and $C_2=-5$ under Lorentzian distribution.}
\end{figure}
The profile of $F_h$, $F_a$, and $F_g$ for our wormhole solutions under Gaussian and Lorentzian sources are shown in Fig. \ref{Fig.13} and \ref{Fig.14}, respectively by assigning the values of the free parameters as we used in the above figures. One can observe from Figs. \ref{Fig.13} and \ref{Fig.14} that the hydrostatic force ($F_h$) is dominating compared to anisotropic ($F_a$) and gravitational ($F_g$) forces, respectively. Here for both the sources, it can be seen that the force $F_h$ takes the positive values while the other forces $F_a$ and $F_g$ are negative, which is clearly define that to hold the system in equilibrium state, the hydrostatic force is balanced by the combined effect of anisotropic and gravitational forces. One may check the Refs. \cite{Rahaman/2015,Abdul/2016} where the authors have been studied deeply on this topic.
\section{Exoticity parameter}
\label{sec6}
\begin{figure}
\centering \epsfig{file=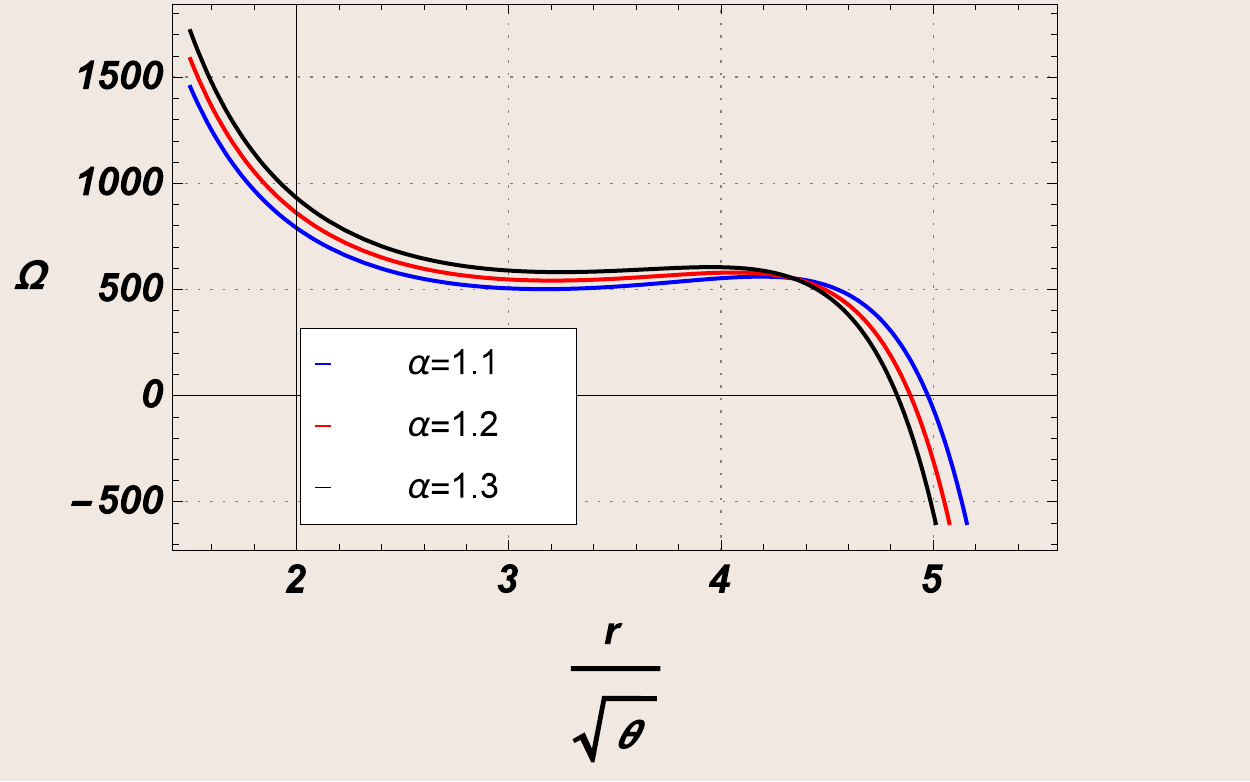, width=.48\linewidth,
height=2.5in}\epsfig{file=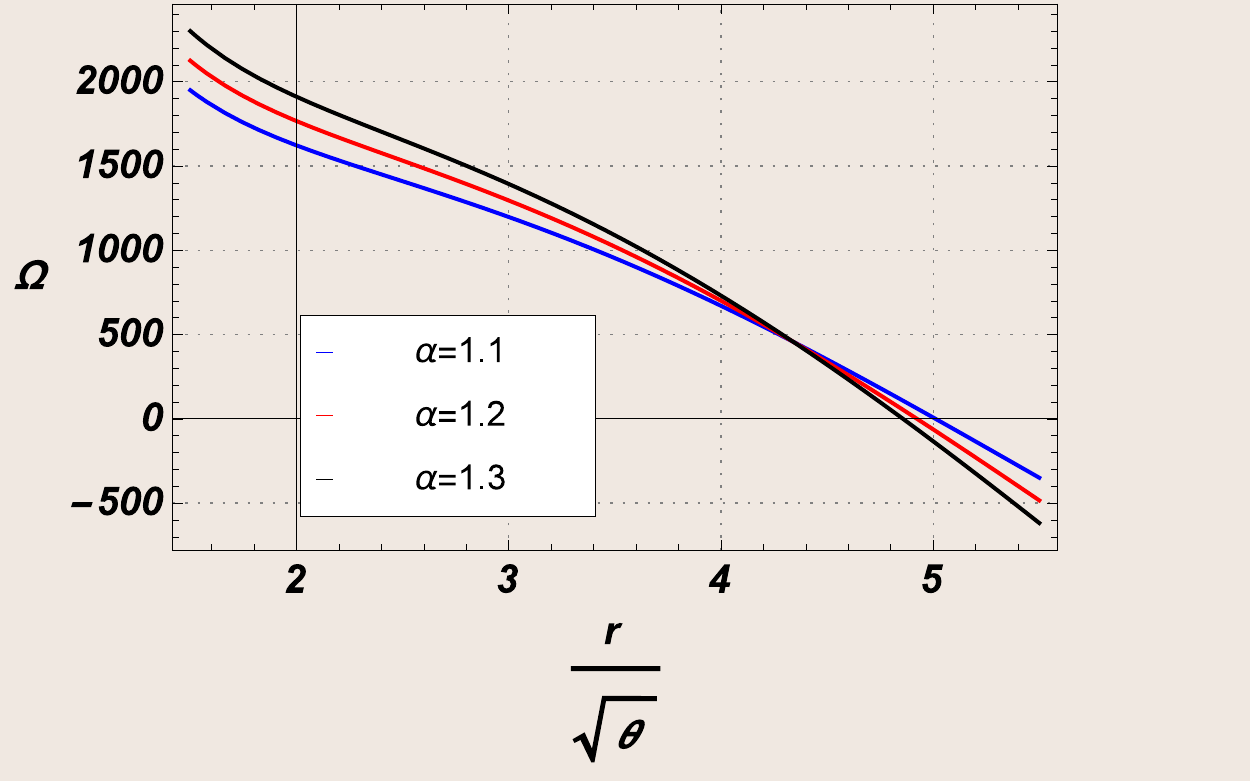, width=.48\linewidth,
height=2.5in}\caption{\label{Fig.15} The profile shows the behavior of exoticity $\Omega$ with respect to $\frac{r}{\sqrt{\theta}}$ for ($left$) Gaussian and ($right$) Lorentzian distributions with fixed parameters $M_1=0.2$, $\beta_1=0.07$, $K_2=1.2$ and $C_1=C_2=-5$.}
\end{figure}

In this section, we shall investigate the presence of the exotic matter at the throat and its vicinity by means of the behavior of the exoticity parameter, i.e., $\Omega$, which is defined as \cite{Visser/1996,Lemon/2003}
\begin{equation}\label{29}
\Omega=-\frac{\rho-p_{r}}{|\rho|},
\end{equation}
Noted that, non-negativity of $\Omega$ ensure the presence of exotic matter at or nearer to the throat of the wormhole. By keeping this concept in mind, using Eqs. \eqref{4a} and \eqref{21} for Gaussian, and Eqs. \eqref{4b} and \eqref{25} for Lorentzian distributions, we have plotted the graphs of exoticity $\Omega$ against the radial coordinate $\frac{r}{\sqrt{\theta}}$ for both distributions in Fig. \ref{Fig.15}. Interestingly, we can observe that exoticity is showing positive behavior at or nearer to the throat of wormhole and after that it is showing negative behavior for both cases. This implies that right from the wormhole's throat to an adequate distance is surrounded by exotic matter. Moreover, we can also conclude that meeting the flare-out condition does not necessarily imply a violation of the NEC in the modified gravity. Readers can also check the Refs. \cite{Morris/1988,Gonzalez/2003,Mubasher/2010} for a detailed discussion.

\section{Concluding remarks}
\label{sec7}
In this manuscript, we have discussed the possibility of the existence of wormhole solutions in the framework of $f(Q)$ gravity or symmetric teleparallel gravity under anisotropy matter sources. The geometry behind this work is inspired by the noncommutativity along with the conformal killing vectors (CKV). The concept of finding a solution with the help of CKVs and non-commutative sources is not a new approach. There are many works available in literature where these concepts have already been used to investigate the existence of black holes and wormholes in various modified theories. Kuhfittig in \cite{Kuhfittig/2018} has discussed different functional forms of $f(R)$ gravity via several shape functions for wormhole under non-commutative geometry. Jamil et al. has used non-commutative geometry to studied wormhole solutions in $f(R)$ gravity in \cite{Jamil/2014}. Also, in the article \cite{Ghosh/2018}, Ghosh studied the existence of Einstein-Gauss-Bonnet black hole inspired by non-commutative geometry. However, this current study provides a novel discussion that such non-commutative distributions along with the conformal symmetry have not been used before in the symmetric teleparallel gravity. In this whole study we assume a particular form of linear model for $f(Q)$ i.e., $f(Q)=\alpha\,Q+\beta$, where $Q$ is the non-metricity scalar which is responsible for the gravitational interaction and $\alpha$ and $\beta$ are free parameters. With the help of this linear functional form, we found that the field equations are extremely complicated, and hence we use conformal symmetry to get the exact analytic solution of field equations. Inspired from the works of Nicolini et al. \cite{Smailagic/2006} and Mehdipour \cite{Mehdipour/2012}, we used non-commutative energy density suggested by them and analyzed the corresponding form of shape functions and energy conditions for obtained wormhole solutions graphically separately. The main features of this current study are highlighted below:
\begin{itemize}
\item[(1)] Under both Gaussian and Lorentzian distribution, we have found that the obtain shape functions are increasing functions in nature with respect to $\frac{r}{\sqrt{\theta}}$ and graphically, it has also shown that at the wormhole's throat, flare out condition hold which is represented in Figs. \ref{Fig.1} and \ref{Fig.7}.
\item[(2)] In the left panel of Figs. \ref{Fig.2} and \ref{Fig.8}, we have shown the behavior of asymptotically flatness condition. We found that due the conformal symmetry assumption, the ratio $\frac{\frac{b}{\sqrt{\theta}}}{\frac{r}{\sqrt{\theta}}}$ approaches to a small positive value as the radial coordinate $\frac{r}{\sqrt{\theta}}$ get larger values. Hence, it can be concluded that the asymptotic flatness behavior of the shape function cannot be achieved in both distributions. As a result, for the redshift function, the wormhole spacetime is not asymptotically flat; hence it will have to be cut off at some radial distance which smoothly joins to an exterior vacuum solution in the standard way.
\item[(3)] Also, one can observe the right panel of Figs. \ref{Fig.2} and \ref{Fig.8} that the values of wormhole throat decrease as increases the values of $\alpha$ in both non-commutative cases.
\item[(4)] In Figs. \ref{Fig.3} and \ref{Fig.9}, it can be seen that, for both cases, energy density shows positively decreasing behavior throughout the spacetime.
\item[(5)] Furthermore, we studied NEC, DEC, and SEC in Figs. \ref{Fig.4}-\ref{Fig.6} and \ref{Fig.10}-\ref{Fig.12}. It is found that for both non-commutative distributions, NEC bounds are violated as $\theta(\rho+p_r)<0$ while the SEC is satisfied. Thus, NEC's violation may confirm the exotic matter at the wormhole's throat, which is a basic requirement for traversable wormhole existence.
\item[(6)] Moreover, the Tolman-Oppenheimer-Volkoff (TOV) equation has been calculated to check the stability of the matter distribution subject to the hydrostatic force $F_h$, anisotropic force $F_a$, and gravitational force $F_g$ due to anisotropic pressure. One can check the Figs. \ref{Fig.13} and \ref{Fig.14} that these forces balanced each other's impact to hold the system in equilibrium and thereby yielding a stable wormhole.
\item[(7)] Lastly, we have used the exoticity parameter to verify the presence of exotic matter at the wormhole's throat. From our analytical analysis presented in Fig. \ref{Fig.15}, it is found that exoticity $\Omega$ is positive at or nearer to the wormhole throat, which implies that the wormhole's throat is filled with exotic matter.
\end{itemize}
Thus, it is very transparent from our analysis that all the necessary conditions relating to a traversable wormhole are fulfilled under the non-commutative geometry and the conformal symmetry. It would also be interesting to explore wormhole solutions in the symmetric teleparallel gravity with the help of different other matter sources into account.

\acknowledgments Z.H. acknowledges Department of Science and Technology (DST), Government of India, New Delhi, for awarding a Junior Research Fellowship (File No. DST/INSPIRE Fellowship/2020/IF190911). PKS acknowledges DST, New Delhi, India for providing facilities through DST-FIST lab, Department of Mathematics, where a part of this work was done. We are very much grateful to the honorable referee and to the editor for the
illuminating suggestions that have significantly improved our work in terms
of research quality, and presentation.


\end{document}